\documentclass[compress,twocolumn]{autart-small-skip}
\usepackage[round]{natbib}
\journal{Elsevier}
\usepackage{graphicx}          % Include this line if your
\usepackage{subfigure}
\usepackage{fancyhdr}
\usepackage{amsmath}
\usepackage{multirow}
\usepackage{amssymb }
\usepackage{color}
\usepackage{graphics} % for pdf,  bitmapped graphics files
\usepackage{graphicx}
\usepackage{amsmath}
\newtheorem{theorem}{\textbf{Theorem}}
\newtheorem{lemma}{\textbf{Lemma}}
\newtheorem{example}{\textbf{Example}}
\newtheorem{corollary}{\textbf{Corollary}}
\newtheorem{remark}{\textbf{Remark}}
\newtheorem{definition}{\textbf{Definition}}

\newtheorem{proposition}{\textbf{Proposition}}

\usepackage{url}
	\newenvironment{proof}{{{\bf Proof:}}}{\hfill $\square$\par}
	
	\usepackage{multirow} %multirow for format of table
	\usepackage{xcolor}
	\usepackage{multirow}
	\usepackage{setspace}
	\usepackage{url}
	\usepackage{subfigure}
	\usepackage{fancyhdr}
	\usepackage{amsmath}
	\usepackage{multirow}
	\usepackage{amssymb }
	\usepackage{color}
	\usepackage{graphics} % for pdf,  bitmapped graphics files
	\usepackage{graphicx}
	\usepackage{subeqnarray}
	\usepackage{cases}
	\usepackage{threeparttable}
	\usepackage{algorithm} %format of the algorithm
	\usepackage{algpseudocode}
	 \normalsize
	
\begin{document}
		
		\begin{frontmatter}
			%\runtitle{Insert a suggested running title}  % Running title for regular
			% papers but only if the title
			% is over 5 words. Running title
			% is not shown in output.
			
			\title{\large{Generic Diagonalizability, Structural Functional Observability and Output Controllability}} % Title, preferably not more
			% than 10 words.
			
			\thanks[footnoteinfo]{This paper was not presented at any IFAC
				meeting. This work was supported in part by the
				National Natural Science Foundation of China under Grant 62373059.}
			
			\author[author1]{Yuan Zhang}\ead{zhangyuan14@bit.edu.cn},
			\author[author2]{Tyrone Fernando}\ead{tyrone.fernando@uwa.edu.au},
			\author[author3]{Mohamed Darouach}\ead{mohamed.darouach@univ-lorraine.fr}
			
			\address[author1]{School of Automation, Beijing Institute of Technology, Beijing, China}
			\address[author2]{Department of Electrical Electronic and Computer Engineering, University of Western Australia, Crawley,Australia}
			\address[author3]{Centre de Recherche en Automatique de Nancy, IUT de Longwy, Universit\'{e} de Lorraine, 54400 Cosnes et Romain, France}
			
			% keyword list or with the
			% help of the Automatica
			% keyword wizard
			%(i.e., systems with generically diagonalizable state matrices)
			
			\begin{abstract} This paper investigates the structural functional observability (SFO) and structural output controllability (SOC) of a class of systems with generically diagonalizable state matrices and explores the associated minimal sensor and actuator placement problems. The verification of SOC and the corresponding sensor and actuator placement problems,  {i.e., the problems of determining the minimum number of outputs and inputs required to achieve SFO and SOC, respectively},  are yet open for general systems, which motivates our focus on a class of systems enabling polynomial-time solutions. In this line, we first define and characterize generically diagonalizable systems, {referring to structured systems for which almost all realizations of the state matrices are diagonalizable}. We then develop computationally efficient criteria for SFO and SOC within the context of generically diagonalizable systems. Our work expands the class of systems amenable to polynomial-time SOC verification.  Thanks to the simplicity of the obtained criteria, we derive closed-form solutions for determining the minimal sensor placement to achieve SFO and the minimal actuator deployment to achieve SOC in such systems, along with efficient weighted maximum matching based and weighted maximum flow based algorithms.
				For more general systems to achieve SFO, an upper bound is given by identifying a non-decreasing property of SFO with respect to a specific class of edge additions, which is shown to be optimal under certain circumstances.

			\end{abstract}

			\begin{keyword}                           % Five to ten keywords,
				Generic diagonalizability, functional observability, output controllability, actuator/sensor selection, graph theory
			\end{keyword}
			
		\end{frontmatter}
	
	{\small{
		
		\section{Introduction} \label{intro-sec}
		
		%	{\color{red}

			%	State observation is one of the fundamental problems in the control community. This is because state observation is prerequisite for various practical applications ranging from observer-based feedback control, fault diagnosis, and attack detection and identification. The problem of designing full-state observer has attracted extensive attention since the pioneering work of Luenberger \cite{1}. On the other hand, for many real-world problems, it is neither desirable nor physically practicable to estimate the entire state vector of a high-dimensional system. Instead, the estimation of speci铿乧 subsets or linear functions of state variables that are of particular interest can serve this need. The problem of designing functional observers to estimate linear combinations of state variables was first proposed by Luenberger \cite{2}, too. Since then, significant achievements been achieved on the functional observer design problems, ranging from linear time-invariant (LTI) systems [10-14], time-varying systems \cite{rotella2012functional}, nonlinear systems [15], and descriptor systems [16].

			State observation is a fundamental issue within the field of control theory, as it forms a crucial prerequisite for a wide range of practical applications, including observer-based feedback control \citep{chen1984linear}, fault diagnosis \citep{Chen1999RobustMF}, and attack detection and identification \citep{F.Pa2013Attack}. The problem of designing a full-state observer has garnered substantial attention since the pioneering work of \cite{luenberger1966observers}. However, in many real-world scenarios, it is neither practical nor desirable to estimate the entire state vector of a high-dimensional system. Instead, the estimation of specific subsets or linear combinations of state variables of particular interest is more applicable. A parallel scenario arises in control of large-scale systems/networks, where it is often impractical to fully control the whole states/nodes, but the control of a prescribed subset or combination of states/nodes may suffice \citep{kreindler1964concepts,gao2014target,van2017distance}.   The problem of designing functional observers to estimate linear combinations of state variables was also initiated by \cite{luenberger1971introduction}. Since then, significant advancements have been made in functional observer designs for various types of systems \citep{darouach2000existence,fernando2010functional,jennings2011existence,darouach2012functional,rotella2015note}.   %, encompassing various system types, including linear time-invariant (LTI) systems \cite{darouach2000existence,fernando2010functional,jennings2011existence,rotella2015note}, time-varying systems \cite{rotella2012functional},  nonlinear systems  \cite{trinh2006partial}, and descriptor systems \cite{darouach2012functional}.
			
			The concept of output controllability, introduced by \cite{kreindler1964concepts} in the 1960s, encompasses the ability to manipulate system's output arbitrarily. By contrast, it was very recently that the concept of functional observability was proposed by \cite{fernando2010functional}. Functional observability represents a fundamental condition for the existence of an observer with arbitrary poles to estimate a linear function of states based on system inputs and outputs. Notice that there is no direct duality between output controllability and functional observability \citep{iudice2019node}. Various algebraic criteria for functional observability have been found \citep{jennings2011existence,functional2022Mohamed,asymptotic2022Mohamed,rotella2015note,zhang2023functional}. These algebraic criteria involve either the system's eigenspace decomposition or observability matrices, emphasizing the necessity of precise parameter values for the system matrices.

			Recently, \cite{montanari2022functional} have expanded the notion of functional observability to structured systems, introducing the concept of structural functional observability (SFO). Similar to structural controllability \citep{C.T.1974Structural}, SFO relies exclusively on the zero-nonzero structure of system matrices, effectively mitigating the influence of parameter uncertainty and rounding-off errors. This distinctive feature of SFO facilitates the development of scalable combinatorial algorithms for both verifying functional observability and optimizing sensor placement \citep{montanari2022functional}. However, it is crucial to highlight that the criterion proposed in \cite{montanari2022functional} is based on a PBH-like rank condition for functional observability, a condition acknowledged as insufficient for general systems recently in \cite{functional2022Mohamed,asymptotic2022Mohamed} and \cite{zhang2023functional}.
        	Indeed, the diagonalizability of state matrices significantly impacts the criteria for functional observability in numerical systems, as recognized in \cite{zhang2023functional}. Specifically, for diagonalizable systems, a PBH-like criterion exists, extending the classical PBH criterion. This phenomenon can be attributed to the fact that eigenvectors of diagonalizable matrices can span the entire space, unlike non-diagonalizable matrices. In the case of non-diagonalizable systems, criteria must depend on either high-degree products of system matrices (similar to the observability matrix) or the observability canonical decomposition \citep{functional2022Mohamed}. The complete criteria for SFO in general systems, as outlined in \cite{zhang2023functional}, are inevitably intricate and require a global understanding of the system structure, such as {the generic dimension of unobservable subspaces}. This complexity poses a significant challenge to the associated sensor placement problems \citep{zhang2023functional}.

			 		 	The study of structural output controllability (SOC), on the other hand, has a relatively long history \citep{murota1990note,Murota_Book,gao2014target,van2017distance,czeizler2018structural,li2020structural}.  Despite this, it remains open whether SOC can be verified in polynomial time by deterministic algorithms \citep{murota1990note}. However, affirmative answers have been found for certain special classes of systems.  For instance,  \cite{gao2014target} developed the $k$-walk theory for structural target controllability of directed tree networks. \cite{li2020structural} provided graph-theoretic characterizations for structural target controllability of undirected networks with symmetric parameters. {Recently, \cite{commault2019functional} proposed a notion termed (structural) functional output controllability, referring to the ability to produce {\emph{any}} smooth output trajectory in ${\mathbb R}^p$ ($p$ is the dimension of system outputs) by designing system inputs. This notion is equivalent to the right-invertibility of system input-output transfer functions \citep{commault2019functional}, which is sufficient for output controllability but not necessary, thus different from the output controllability discussed in this paper.} {Another research line is strong structural output controllability, which requires all system realizations obtained by assigning nonzero values to the indeterminate parameters to be output controllable \citep{van2017distance,shali2021properties}. Some sufficient conditions for this property can be found in \cite{shali2021properties}.}
%in ${\mathbb R}^p$ ($p$ is the dimension of system outputs)			 	

In this paper, we focus on the SFO and SOC, along with the associated minimal sensor and actuator placement problems, {i.e., the problems of determining the minimum number of outputs (sensors) and inputs (actuators) required to achieve SFO and SOC, respectively}, within a class of systems with generically diagonalizable state matrices (termed generically diagonalizable systems).
			 	We begin by providing explicit characterizations for generically diagonalizable matrices, {referring to structured matrices for which almost all realizations are diagonalizable}. We then demonstrate that within the domain of generically diagonalizable systems, the aforementioned verification and optimization problems can be efficiently addressed in polynomial time. We also extend our analysis to encompass general cases. Our main contributions are as follows:
			 	
			 	\begin{itemize}
			 		\item 	Firstly, we define generically diagonalizable matrices (and structurally diagonalizable graphs) and propose graph-theoretic characterizations for them. We establish that a graph is structurally diagonalizable if and only if each subgraph induced by every subset of strongly-connected components (SCCs) of this graph is so. We also propose a weighted maximum matching based algorithm to identify generic diagonalizability.
			 		\item  Secondly, we develop simplified criteria for SFO in generically diagonalizable systems. These criteria are significantly simpler compared to those for general systems given in \cite{zhang2023functional}, as they no longer necessitate the system global information, such as the {generic dimension of unobservable subspaces}, {which is beneficial for solving the subsequent minimal sensor placement problem.}  Notably, our criteria affirm the applicability of the criterion in \cite{montanari2022functional} to the class of generically diagonalizable systems introduced in this paper, despite its derivation from a different SFO definition.
			 		\item  Thirdly, we identify a class of systems for which the SOC can be verified in polynomial time and highlight that generically diagonalizable systems fall into it.
			 		\item   Lastly, leveraging our established criteria, we present a closed-form solution and a weighted maximum matching based algorithm for the minimal sensor placement problem for achieving SFO in generically diagonalizable systems. For more general systems, we identify a non-decreasing property of SFO with respect to (w.r.t.) a specific class of edge additions and propose two algorithms to obtain an upper bound, proven optimal under certain circumstances. We also propose a weighted maximum flow algorithm to determine the minimal actuators needed for SOC in generically diagonalizable systems. Remarkably, for general systems, both of these problems remain open.
			 	\end{itemize}
			 	
%			 	 We first define and characterize generically diagonalizable matrices and then show that within the context of generically diagonalizable systems, the aforementioned verification and optimization problems can be solved in polynomial time. We also consider the general cases. Our main contributions are as follows:

			%    These criteria indicate that the criterion provided in \cite{montanari2022functional}, while derived from a different SFO definition than ours and not universally sufficient, is indeed applicable to the class of generically diagonalizable systems introduced in this paper.

%	           The study of generic diagonalizable matrices holds independent interest. For example, the diagonalizablity of state matrices can also dramatically affect criteria in relation to other system properties, such as Lyapunov stability of modes with zero real parts \cite{chen1984linear} and the controllability of networked high-order systems \cite{zhang2021structural,xue2021modal}. Besides,  diagonalizable systems frequently arise in practical applications, such as systems with symmetric state matrices, non-zero random diagonal entries \cite{park2017design}, or simple spectra \cite{tao2017random}.
	
	           The study of generically diagonalizable matrices holds independent interest, impacting criteria related to various system properties, such as Lyapunov stability of modes with zero real parts \citep{chen1984linear} and the controllability of networked high-order systems \citep{xue2021modal,zhang2021structural}. Furthermore, diagonalizable systems find frequent application in practical scenarios, including systems with symmetric state matrices, non-zero random diagonal entries or simple spectra \citep{tao2017random}, and distributed estimation \citep{park2017design}.

		This paper is organized as follows.  Section \ref{pre-sec} presents some basic preliminaries. Section \ref{section_prob} reviews the concepts of SFO and SOC. Characterizations of generically diagonalizable matrices are provided in Section \ref{sec-characterize}. Computationally efficient criteria for SFO and SOC of generically diagonalizable systems are given respectively in Section \ref{sec-sfo-diagonal} and Section \ref{soc-section}. Section \ref{sec-sensor} is devoted to the assocaited minimal sensor and actuator placement problems. Section \ref{sec-conclude} concludes this paper.

			{\bf Notations:} {Define the set $[n]\doteq \{1,2,\ldots,n\}$ for any integer $n\ge 1$. Denote the set of non-negative real numbers by ${\mathbb R}_{\ge 0}$. The composite matrix stacked by $X_1,...,X_n$ is denoted as ${\bf col}\{X_i|_{i=1}^n\}$, where $X_i$ represents the $i$th row block. When $n$ is small, ${\bf col}\{X_1,...,X_n\}$ can also be represented as $[X_1;...;X_n]$. The identity matrix of dimension $n$ is denoted as $I_n$, and the subscript $n$ can be omitted if inferred from the context. For an $m\times n$ matrix $M$ and two sets $S_1\subseteq [m]$ and $S_2\subseteq [n]$, the sub-matrix of $M$ formed by rows indexed by $S_1$ and columns indexed by $S_2$ is denoted as ${M}(S_1,S_2)$. When $S_1=[m]$ ($S_2=[n]$), $M(S_1,S_2)$ is also denoted as $M(:,S_2)$ ($M(S_1,:)$). The number of rows of a matrix $M$ is denoted by ${\rm row}(M)$.}
			
			\section{Preliminaries} \label{pre-sec}
			This section introduces preliminaries in graph theory and structured system theory \citep{generic,Ramos2022AnOO}.

			{\bf Graph-theoretic terminologies:} A directed graph (digraph) is represented by ${\mathcal G}=(V,E)$, where $V$ is the vertex set and $E\subseteq V\times V$ is the edge set. A subgraph ${\mathcal G}_s=(V_s,E_s)$ of ${\mathcal G}$ is a graph such that $V_s\subseteq V$ and $E_s\subseteq E$.  It is called a subgraph induced by $V_s$ if $E_s=(V_s\times V_s) \cap E$. We say that ${\mathcal G}_s=(V_s,E_s)$ spans ${\mathcal G}=(V,E)$ if $V_s=V$, and that ${\mathcal G}_s$ {\emph{covers}} a subset $V_s'\subseteq V$ if $V_s'\subseteq V_s$. A path $P$ from vertex $i_1$ to vertex $i_k$ in ${\mathcal G}$ is a sequence of edges $(i_1,i_2)$, $(i_2,i_3)$,...,$(i_{k-1},i_k)$ with $(i_j,i_{j+1})\in E$, $j=1,...,k-1$. Denote such a path by $P=(i_1,i_2,...,i_k)$. If there is a path from $i_1$ to $i_k$, we say $i_k$ is {\emph{reachable}} from $i_1$. A path with no repeated vertices is referred to as a simple path. The length of a path is the number of edges it contains. A vertex can be seen as a path with zero length. %If the only repeated vertices in path $P$ are the starting and ending vertices ($i_1$ and $i_k$), then $P$ is called a {\emph{cycle}} For two digraphs ${\mathcal G}_1=(V_1,E_1)$ and ${\mathcal G}_2=(V_2,E_2)$, ${\mathcal G}_1\cup {\mathcal G}_2$ denotes the digraph $(V_1\cup V_2, E_1\cup E_2)$.
			
			A bipartite graph ${\mathcal B}=(V,E)$ is a graph where the vertex set $V$ can be partitioned into two disjoint sets $V_L$ and $V_R$ (called parts of $V$) such that $E\subseteq V_R\times V_L$. In this case, ${\mathcal B}$ is also denoted by ${\mathcal B}=(V_L, V_R, E)$.\footnote{Although edges in a bipartite graph are unoriented, in this paper it is convenient to consider that they have a direction from $V_R$ to $V_L$.} A matching of ${\mathcal B}$ is a set of edges ${\mathcal M}\subseteq E$ in which no two edges share a common end vertex. The size of ${\mathcal M}$, denoted as $|{\mathcal M}|$, is the number of edges in ${\mathcal M}$. A vertex $v$ is considered {\emph{left matched}} (resp. right matched) by ${\mathcal M}$ if $v\in V_L$ (resp. $v\in V_R$) is incident to an edge in ${\mathcal M}$.  A {\emph{perfect matching}} is a matching that left matches $V_L$ and right matches $V_R$.  A maximum matching of ${\mathcal B}$ refers to a matching with the largest size among all possible matchings of ${\mathcal B}$. If each edge $e\in E$ is assigned a non-negative cost, the weight of a matching is the sum of the costs of its edges. The weighted maximum matching problem seeks the maximum (or minimum) weight of a maximum matching among all maximum matchings. This problem can be solved in polynomial time \citep{Ahuja1993NetworkFT}. A {\emph{strongly connected component}} (SCC) of a directed graph ${\mathcal G}=(V,E)$ is a subgraph ${\mathcal G}_s$ of ${\mathcal G}$ that satisfies the following conditions: for any pair of vertices in ${\mathcal G}_s$, there exists a path from either vertex to the other, and no additional vertices can be included in ${\mathcal G}_s$ without violating the previous property.

			{\bf{ Structured system theory:}} 	Consider a linear-time invariant (LTI) system described by	
			\begin{subequations}  \label{pre-system}
				\begin{align}
					\dot x(t)&=Ax(t)+Bu(t), \label{pre-state-eq}\\
					y(t)&=Cx(t), \label{pre-lumped-output}
				\end{align}
			\end{subequations}
			where $x(t)\in {\mathbb R}^{n}$ is the state vector, $u(t)\in{\mathbb R}^{m} $ is the input vector, and $y(t)\in {\mathbb R}^{p}$ is the output vector.
			
			{A structured matrix is a matrix whose entries are either fixed zero or free parameters that can take arbitrary real values independently (i.e., no parameter dependence exists among those parameters).} We denote the set of $n_1\times n_2$ structured matrices by $\{0,*\}^{n_1\times n_2}$,  where $*$ represents the free parameters, and $0$ the fixed zero entries.
			For a structured matrix $\bar M\in \{0,*\}^{n_1\times n_2}$, let $n_{\bar M}$ denote the number of $*$ entries in $\bar M$. A realization is obtained by assigning some specific values to the $*$ entries. We use the vector gathering values for $*$ entries $\theta\in {\mathbb R}^{n_{\bar M}}$ to denote a realization of $\bar M$ (with some abuse of notation, since they are in one-to-one correspondence). For a subset ${\mathbb V}\subseteq {\mathbb R}^{n_{\bar M}}$, $M\in {\mathbb V}$ means that $M$ is a realization of $\bar M$ corresponding to some $\theta\in {\mathbb V}$.

{An algebraic variety in ${\mathbb R}^d$ is the set of real-valued solutions of a system of polynomial equations in $d$ variables. A proper variety in ${\mathbb R}^d$ is an algebraic variety that is not the entire space ${\mathbb R}^d$ (i.e., it is defined by at least one non-zero polynomial equation).}	The {\emph{generic rank}} of a structured matrix $\bar M$, given by ${\rm grank}\, \bar M$, is the maximum rank the realizations of $\bar M$ can achieve as a function of its free parameters.
{Furthermore, let the entries $P_{ij}$ of a matrix $P$ be polynomials in $d$ free parameters (for example, $P$ can be the product of a sequence of structured matrices in $d$ free parameters).
 Its {\emph{generic rank}} ${\rm grank}\, P$ is the maximum rank this matrix can achieve as a function of the $d$ free parameters in $P$.} Here, the generic rank also equals the rank that this matrix can achieve for almost all choices of parameter values (i.e., all except for some proper variety) in the parameter space ${\mathbb R}^d$ {\cite[page 38]{Murota_Book}}.
			For a structured matrix $\bar M\in \{0,*\}^{n_1\times n_2}$, the bipartite graph associated with $\bar M$, given by ${\mathcal B}(\bar M)$, is defined as $(X_L,X_R,E_{RL})$, where $X_L$, $X_R$ correspond to respectively the {\emph{rows}} and {\emph{columns}} of $\bar M$, and $E_{RL}=\{(i,j): \bar M_{ji}\ne 0, i\in X_R, j\in X_L\}$. {It follows from \citet[Props. 2.1.12 and 2.2.25]{Murota_Book} that ${\rm grank}\,\bar M$ equals the size of a maximum matching of ${\mathcal B}(\bar M)$.}
			
			Consider a triple $(\bar A, \bar B, \bar C)$ with $\bar A\in \{0,*\}^{n\times n}$, $\bar B\in \{0,*\}^{n\times m}$, and $\bar C\in \{0,*\}^{p\times n}$. We define the {\emph{digraph}} ${\mathcal G}(\bar A, \bar B, \bar C)=(X\cup U\cup Y, E_{UX}\cup E_{XX}\cup E_{XY})$ as follows: the state vertices $X\!=\!\{x_1,...,x_n\}$, the input vertices $U=\{u_1,...,u_m\}$, the output vertices $Y=\{y_1,...,y_p\}$, and the edges $E_{XX}=\{(x_i,x_j): \bar A_{ji}\ne 0\}$, $E_{UX}=\{(u_i,x_j):\bar B_{ji}\ne 0\}$, and $E_{XY}=\{(x_i,y_j): \bar C_{ji}\ne 0\}$.  Let ${\mathcal G}(\bar A)$, ${\mathcal G}(\bar A, \bar B)$, and ${\mathcal G}(\bar A, \bar C)$ be the subgraphs of ${\mathcal G}(\bar A, \bar B, \bar C)$ induced by $X$, $X\cup U$, and $X\cup Y$, respectively. Define a bipartite graph ${\mathcal B}(\bar A)=(X,X,E_{XX})$, in which with some abuse of notations, we allow two vertices in different parts to have the same label. In this way, for a matching $\mathcal M$ in ${\mathcal B}(\bar A)$, $(X,{\mathcal M})$ is a subgraph of ${\mathcal G}(\bar A)$ with vertex set $X$ and edge set ${\mathcal M}\subseteq E_{XX}$.
			  A state vertex $x_i\in X$ is considered {\emph{output-reachable}} (resp. {\emph{input-reachable}}) if $x_i$ is reachable to some output vertex $y_j\in Y$ in ${\mathcal G}(\bar A, \bar C)$ (resp. reachable from some input vertex $u_j\in U$ in ${\mathcal G}(\bar A, \bar B)$). A state vertex $x_i\in X$ has a self-loop if $(x_i,x_i)\in E_{XX}$.  An {\emph{output stem}} is a simple path from a state vertex to an output vertex in ${\mathcal G}(\bar A, \bar C)$. Similarly, an {\emph{input stem}} is a simple path from an input vertex to a state vertex in ${\mathcal G}(\bar A, \bar B)$.  An {\emph{output cactus configuration}} of ${\mathcal G}(\bar A, \bar C)$  is a {\emph{subgraph}} consisting of a collection of vertex-disjoint output stems and cycles, where each vertex in the cycles is output-reachable in ${\mathcal G}(\bar A, \bar C)$ (notice that the paths making each cycle output-reachable are not included in a cactus configuration). Similarly, we can define {\emph{input cactus configuration}} by changing ``output stems'' to ``input stems'' in ${\mathcal G}(\bar A, \bar B)$. The size of an input/output cactus configuration is the {\emph{number of state vertices}} it covers.

% is a subgraph consisting of a collection of vertex-disjoint input stems and cycles, where each vertex in the cycles is input-reachable in ${\mathcal G}(\bar A, \bar B)$.
			
			   A pair $(\bar A,\bar C)$ (resp. $(\bar A,\bar B)$) is said to be {\emph{structurally observable}} ({\emph{structurally controllable}}), if there is a realization $(A,C)$ of $(\bar A, \bar C)$ that is observable (resp. a realization $(A,B)$ of $(\bar A, \bar B)$ that is controllable) \citep{C.T.1974Structural}.
			It is known that the pair $(\bar A, \bar C)$ is structurally observable (resp. $(\bar A, \bar B)$ is structurally controllable), if and only if the whole state vertex set $X$ is covered by an output cactus configuration in ${\mathcal G}(\bar A, \bar C)$ (resp. input cactus configuration in ${\mathcal G}(\bar A, \bar B)$). {See \citet[Theo 1]{commault2008observability} and \citet[Theo 1]{generic} for other equivalent criteria for structural observability. } %.  $X$ is covered by a cactus configuration in  ${\mathcal G}(\bar A, \bar C)$
% {i) every $x_i\in X$ is output reachable in ${\mathcal G}(\bar A, \bar C)$, and ii) ${\rm grank}\,[\bar A; \bar C]=n$ \citet[Theo 14.2]{K.J.1988Multivariable}.} 			
			% A set $X_s\subseteq X$ is said to be {\emph{covered by a cactus configuration}} in ${\mathcal G}(\bar A, \bar C)$ if the following conditions hold: 1) every $x_i\in X_s$ is output-reachable, and 2) $X_s$ can be covered by a collection of vertex-disjoint cycles and output stems.

					\section{SFO and SOC} \label{section_prob}
		Consider an LTI system described by	
		\begin{subequations}  \label{lumped-system}
			\begin{align}
				\dot x(t)&=Ax(t)+Bu(t), \label{state-eq}\\
				y(t)&=Cx(t), \label{lumped-output} \\
				z(t)&=Fx(t),   \label{functional-info}
			\end{align}
		\end{subequations}
		where $x(t)\in {\mathbb R}^{n}$ is the state vector, $u(t)\in{\mathbb R}^{m} $ is the input vector, $y(t)\in {\mathbb R}^{p}$ is the output vector, and $z(t)\in {\mathbb R}^{r}$ is the functional of states to be estimated. Accordingly, matrices $A\in {\mathbb R}^{n\times n}$, $B\in {\mathbb R}^{n\times m}$, $C\in {\mathbb R}^{p\times n}$, and $F\in {\mathbb R}^{r\times n}$. Throughout this paper,  $O(A,C)={\bf col}\{C,CA,...,CA^{n-1}\}$ denotes the observability matrix of the pair $(A,C)$, and $Q(A,B)=[B,AB,\cdots, A^{n-1}B]$ denotes the controllability matrix of the pair $(A,B)$.
		
		\begin{definition}[\citeauthor{Modern_Control_Ogata},\citeyear{Modern_Control_Ogata}]
			System (\ref{lumped-system}) (or the triple $(A,B,C)$) is output controllable, if for any initial output $y_0\in {\mathbb R}^p$ and any final output $y_f\in {\mathbb R}^p$,  there exist a finite time $T$ and control input $u(t): [0,T]\to {\mathbb R}^m$ such that $y(0)=y_0$ and $y(T)=y_t$.
		\end{definition}
		
		\begin{definition}[\citeauthor{fernando2010functional2},\citeyear{fernando2010functional2}]\label{functional-def} System (\ref{lumped-system}) (or the triple $(A,C,F)$) is said to be functionally observable, if for any initial state $x(0)$ and input $u(t)$, there exists a finite time $T$ such that the value of $Fx(0)$ can be uniquely determined from the outputs $y(t)$ and inputs $u(t)$, $0\le t \le T$.
		\end{definition}
		
		It is easy to see that when $F=I_n$, functional observability collapses to the conventional observability \citep{chen1984linear}. The following lemma presents a PBH-like criterion for the functional observability of a class of systems with diagonalizable state matrices $A$, which are called {\emph{diagonalizable systems}} throughout.  {It is worth mentioning that this condition is necessary, but generically insufficient, for non-diagonalizable systems \citep{zhang2023functional}}.
		
		\begin{lemma}({\bf \citeauthor{functional2022Mohamed},\citeyear{functional2022Mohamed};\citeauthor{zhang2023functional},\citeyear{zhang2023functional}}) \label{functional-theorem-diagonal}
			Suppose $A$ is diagonalizable. The triple $(A,C,F)$ is functionally observable if and only if
			{\begin{equation}\label{rank-criterion-observ} {\rm rank}[A-\lambda I_n;C;F]={\rm rank}[A-\lambda I_n ; C], \forall \lambda\in {\mathbb C}.\end{equation}}
		\end{lemma}
		
		Let $\bar A\in \{0,*\}^{n\times n}$, $\bar B\in \{0,*\}^{n\times m}$, $\bar C\in \{0,*\}^{p\times n}$, and $\bar F\in \{0,*\}^{r\times n}$ be structured matrices specifying the zero-nonzero structure of $A, B, C$, and $F$, respectively. That is, $\bar M_{ij}=0$ implies $M_{ij}= 0$  for $M=A,B,C$ or $F$. The triple $(\bar A, \bar B, \bar C)$ is SOC, if there is a realization of $(\bar A, \bar B, \bar C)$ that is output controllable {\citep{murota1990note,czeizler2018structural}}. By contrast, SFO is defined differently, shown as follows.
		
		\begin{definition}[\citeauthor{zhang2023functional},\citeyear{zhang2023functional}]\label{def-structural-functional}  The triple $(\bar A, \bar C, \bar F)$ is SFO, if there is a proper variety ${\mathbb V}\subseteq {\mathbb R}^{n_{\bar A}}\times {\mathbb R}^{n_{\bar C}}\times {\mathbb R}^{n_{\bar F}}$ (which is a variety in ${\mathbb R}^{n_{\bar A}+n_{\bar B}+n_{\bar C}}$), such that all realizations $(A, C, F)\in {\mathbb R}^{n_{\bar A}}\times {\mathbb R}^{n_{\bar C}}\times {\mathbb R}^{n_{\bar F}} \backslash {\mathbb V}$ are functionally observable.
		\end{definition}		
		
		In other words, $(\bar A, \bar C, \bar F)$ is SFO if {\emph{almost all}} realizations of $(\bar A, \bar C, \bar F)$ are functionally observable. From \cite{zhang2023functional},
		if $(\bar A, \bar C, \bar F)$ is not SFO, then almost all its realizations are not functionally observable, except for possibly a proper variety in ${\mathbb R}^{n_{\bar A}}\times {\mathbb R}^{n_{\bar C}}\times {\mathbb R}^{n_{\bar F}}$ \citep{zhang2023functional}. It is worth noting that \cite{montanari2022functional} has defined SFO in the same way as structural observability, i.e., as the property of enabling a functionally observable realization. As demonstrated in \cite{zhang2023functional} and also illustrated in the following example, there is a key distinction between SFO and structural observability. In structural observability, the existence of an observable realization implies that almost all realizations are observable.  However, the existence of a
		functionally observable realization cannot guarantee that almost all realizations are functionally observable. %; see the following example.

%It is important to note that even for generically diagonalizable systems satisfying assumptions 1) and 2) in the preceding remark, the existence of a functionally observable realization does not imply that almost all realizations of a structured system are functionally observable. This becomes evident from the example below.
{	\begin{example} \label{example-counter}
			Consider a structured system with
			{\small$\bar A=\left[\begin{array}{cccc}0 & 0 & 0 & * \\ 0 & 0 & 0 & * \\ 0 & 0 & 0 & * \\ 0 & 0 & 0 & * \end{array}\right],\begin{array}{c}
					\bar C=\left[\begin{array}{cccc} * & * & * & 0 \\ * & * & * & 0 \\0 & 0 & 0 & * \end{array}\right],\\
					\bar F=\left[\begin{array}{cccc} * & 0 & 0 & 0 \end{array} \right].
				\end{array}$}\\There is a realization of $(\bar A, \bar C, \bar F)$ as\\
			{\small
				$A=\left[\begin{array}{cccc}0 & 0 & 0 & 1 \\ 0 & 0 & 0 & 1 \\ 0 & 0 & 0 & 1 \\ 0 & 0 & 0 & 1 \end{array}\right],\begin{array}{c}
					C=\left[\begin{array}{cccc} 1 & 1 & 1 & 0 \\ 2 & 1 & 1 & 0 \\0 & 0 & 0 & 1 \end{array}\right],\\
					F=\left[\begin{array}{cccc} 1 & 0 & 0 & 0 \end{array} \right].
				\end{array}$\\}
			It can be verified that $A$ is diagonalizable and $(A,C,F)$ is functionally observable via Lemma \ref{functional-theorem-diagonal}. However, it turns out that ${\rm grank}\,[\bar A;\bar C;\bar F]=4\ne {\rm grank}\,[\bar A;\bar C]=3$. This means, for almost all realizations $(A,C,F)\in {\mathbb R}^{n_{\bar A}}\times {\mathbb R}^{n_{\bar C}}\times {\mathbb R}^{n_{\bar F}}$, ${\rm rank}[A;C;F]=4\ne {\rm rank}[A;C]=3$, leading to the functional unobservability of $(A,C,F)$ (see the comment above Lemma \ref{functional-theorem-diagonal}~).
%Theorem \ref{main-theorem} on $(\bar A, \bar C, \bar F)$ yields that almost all realizations of $(\bar A, \bar C, \bar F)$ are not functionally observable.  %Theorem \ref{generic-diagonal-theorem} gives that $\bar A$ is generically diagonalizable, and subsequently %(in fact, $\bar A$ is generically diagonalizable, a notion to be defined in the next section)
		\end{example}	}
		
		\begin{lemma}[Prop 2, \citeauthor{zhang2023functional},\citeyear{zhang2023functional}]\label{characterization-pro}
		{	The following statements are equivalent:
			\begin{itemize}
                \item[(i)] The triple $(\bar A, \bar C, \bar F)$ is SFO.
				\item[(ii)] ${\rm grank}\,([O(\bar A, \bar C);\bar F])={\rm grank}\,O(\bar A, \bar C)$.
				\item[(iii)]  ${\rm grank}\,([O(\bar A, \bar C);O(\bar A, \bar F)])={\rm grank}\,O(\bar A, \bar C)$.
			\end{itemize}}
		\end{lemma}
%\footnote{${\rm grank}\,O(\bar A, \bar C)$ (similar for other matrices) is the maximum rank that $O(\bar A, \bar C)$ can achieve as a function of free parameters in $\bar A, \bar C$.}
%conditions are equivalent necessary and sufficient conditions for SFO of $(\bar A, \bar C, \bar F)$\footnote{${\rm grank}\,O(\bar A, \bar C)$ (similar for other matrices) is the maximum rank that $O(\bar A, \bar C)$ can achieve as a function of free parameters in $\bar A, \bar C$.} %(which also equals the rank of $O(A, C)$ for almost all realizations $(A,C)$ of $(\bar A, \bar C)$)
		
		It is known that $(\bar A,\bar B,\bar C)$ is SOC \citep{murota1990note}, if and only if $${\rm grank}\, \bar CQ(\bar A,\bar B)=p.$$
		
		The primary goal of this paper is to characterize SFO, SOC, and investigate the associated minimal sensor and actuator placement problems within the class of diagonalizable systems, demonstrating their polynomial time solvability. {Specially, the minimal sensor and actuator placement problems aim to minimize the number of sensors (rows of $\bar C$) and actuators (columns of $\bar B$) required to achieve SFO and SOC, respectively, which will be formally defined in Section \ref{sec-sensor}.} To this end, we begin by defining  `diagonalizable' structured matrices/systems and characterizing this class of matrices, referred to as generically diagonalizable matrices.   Then, we develop computationally efficient criteria for SFO and SOC in generically diagonalizable systems. Finally, we provide closed-form solutions for the {aforementioned} minimal sensor and actuator placement problems in such systems, as well as an upper bound for general systems. % Following this, we give closed-form solutions to the associated minimal sensor and actuator placement problems for such systems, and an upper bound for general systems.%, To this end, we first define what it means for a structured matrix to be `diagonalizable' and characterize this class of matrices (termed generically diagonalizable matrices).

				\section{Characterizations of generically diagonalizable matrices} \label{sec-characterize}
			In this section, we define and characterize generically diagonalizable matrices. We also relate generic diagonalizability to SCCs and provide an efficient algorithm to verify generic diagonalizability. Our characterizations can be used to identify generically diagonalizable systems studied in the subsequent sections.
			\begin{definition}[{Generic diagonalizability}]\label{def-generic-diag}
				A structured matrix $\bar A\in \{0,*\}^{n\times n}$ is called generically diagonalizable, if almost all realizations of $\bar A$ are diagonalizable, i.e., there exists a proper variety ${\mathbb V} \subseteq {\mathbb R}^{n_{\bar A}}$ such that all realizations $A\in {\mathbb R}^{n_{\bar A}}\backslash {\mathbb V}$ are diagonalizable.   System (\ref{lumped-system}) is said to be generically diagonalizable, if $\bar A$ is generically diagonalizable.
			\end{definition}

			Notice that the existence of a diagonalizable realization does not imply generic diagonalizability (for example, $0_{n\times n}$ is a diagonalizable realization of any $\bar A\in \{0,*\}^{n\times n}$).	%The following two lemmas are useful for our further derivations.

			\begin{lemma}[\citeauthor{hosoe1979irreducibility},\citeyear{hosoe1979irreducibility}]  \label{hosoe_irreducibility}
				Given a structured pair $(\bar A, \bar C)$, there exists a proper variety ${\mathbb V}\subseteq {\mathbb R}^{n_{\bar A}}\times {\mathbb R}^{n_{\bar C}}$, such that for all realizations $(A,C)\in {\mathbb R}^{n_{\bar A}}\times {\mathbb R}^{n_{\bar C}}\backslash {\mathbb V}$, 1) every nonzero eigenvalue of $A$ is simple (i.e., with algebraic multiplicity one), and 2) if all state vertices in ${\mathcal G}(\bar A, \bar C)$ are output-reachable, then each nonzero eigenvalue (mode) of $A$ is observable, i.e., ${\rm rank}\,[A-\lambda I_n; C]=n$ for each nonzero eigenvalue $\lambda$ of $A$. %Simply speaking, for almost all realizations $(A,C)$ of $(\bar A, \bar C)$, every nonzero eigenvalue of $A$ is simple, and if every $x\in X$ is output-reachable, then every nonzero mode of $A$ is observable.
			\end{lemma}
			
			\begin{lemma}{\bf(Theo A2.1,\citeauthor{K.J.1988Multivariable},\citeyear{K.J.1988Multivariable}; Lem 3, \citeauthor{S.Pe2016A}, \citeyear{S.Pe2016A})} \label{matching-decomposition}
				Given $\bar A\in \{0,*\}^{n\times n}$, let ${\mathcal M}$ be any maximum matching of ${\mathcal B}(\bar A)$. Then, $(X,{\mathcal M})$ is a union of disjoint cycles and paths, and the number of paths equals $n-|{\mathcal M}|$ (an isolated vertex is regarded as a path with length zero), { recalling that $(X,{\mathcal M})$ is a digraph with vertex set $X$ and edge set ${\mathcal M}\subseteq E_{XX}$ that spans ${\mathcal G}(\bar A)$}. %, each of which starts from a right-unmatched vertex of ${\mathcal B}(\bar A)$ and ends at a left-unmatched one
			\end{lemma}
			
			\begin{theorem} \label{generic-diagonal-theorem}
				Given a structured matrix $\bar A\in \{0,*\}^{n\times n}$, the following statements are equivalent:
				\begin{itemize}
					\item[(a)] $\bar A$ is generically diagonalizable.
					\item[(b)] ${\rm grank}\,\bar A=v(\bar A)$,   where $v(\bar A)$ is the maximum number of vertices that are covered by a collection of (vertex) disjoint cycles in ${\mathcal G}(\bar A)$.
					%{\emph{the \ maximum \ number\ of}} \ {\emph{vertices \ covered \ by \ disjoint \ cycles \ in }} \ {\mathcal G}(\bar A)$
					%	$n-{\rm grank}\,\bar A= {\emph{the \ minimum \ number\ of}} \ \lambda-{\emph{edges \ in \ a \ n-matching \ of }} \ {\mathcal B}(\bar A-\lambda I)$.
					%  $ {\rm grank}\,\bar A= {\emph{the \ minimum \ number\ of}} \ \lambda-{\emph{edges \ in \ all \ n-matchings\ of }} \ {\mathcal B}(\bar A-\lambda I)$.
					\item[(c)] There is a maximum matching ${\mathcal M}$ of ${\mathcal B}(\bar A)$ such that the digraph ${\mathcal G}({\mathcal M})\doteq(X,{\mathcal M})$ is a union of disjoint cycles and isolated vertices.
					%				\item[(d)] There exists no permutation matrix $P$ such that
					%				\begin{equation}\label{diagonal-standard-form}
						%					P\bar A P^{-1} = \left[\begin{array}{@{}c|c@{}}
							%						\begin{matrix}
								%							0 & ? & \cdots & ? \\
								%							0 & 0 & \cdots & ?  \\
								%							\vdots & \vdots & \ddots & \vdots \\
								%							0 & 0 & \cdots & 0
								%						\end{matrix}
							%						& \begin{matrix}
								%							0 & \cdots & 0  \\
								%							0 & \cdots & 0  \\
								%							\vdots & \vdots & \vdots \\
								%							0 & \cdots & 0
								%						\end{matrix} \\
							%						\hline
							%						\bar A_{21} &
							%						\bar A_{22}
							%					\end{array}\right],\end{equation}where $?$ entries can be either $*$ or $0$, but at least one of which is $*$, and $\bar A_{22}$ is of dimension $n_2\times n_2$, $0\le n_2\le n-1$.
					%				
					%					$P\bar A P^{-1}=\left[\begin{array}{cccc}
						%					D_1 & 0 & \cdots & 0 \\
						%					D_{12} & D_2 & \cdots & 0 \\
						%					\vdots & \vdots & \ddots & \vdots \\
						%					D_{r1} & D_{r2} & \cdots & D_r
						%				\end{array}\right]$
					%				
					%				$D_i=\left[\begin{array}{cccc}
						%					0 & ? & \cdots & ? \\
						%					0 & 0 & \cdots & ?  \\
						%					\vdots & \vdots & \ddots & \vdots \\
						%					0 & 0 & \cdots & 0
						%				\end{array}\right]$
				\end{itemize}
				
			\end{theorem}
			
			\begin{proof}
				(a) $\Leftrightarrow$ (b): Let $A$ be a realization of $\bar A$. Expand $\det (\lambda I-A)$ as
				$$\det(\lambda I-A)= \lambda^n + a_1\lambda^{n-1}+\cdots + a_n,$$
				where, by the definition of determinant, the coefficients $$a_k=(-1)^k\sum_{J=\{j_1,...,j_k\}\subseteq [n],j_1<\cdots<j_k} \det A(J,J), \forall k\in [n].$$
				Let $\mu(\bar A)=\max \{|J|: {\rm grank}\, \bar A(J,J)=|J|, J\subseteq [n]\}$. Using Lemma \ref{matching-decomposition} on ${\mathcal B}(\bar A(J,J))$, $J\subseteq [n]$, we have $\mu(\bar A)=v(\bar A)$. It then follows that $a_k=0$ if $k> v(\bar A)$. Therefore, $\det(\lambda I-A)$ can be expressed as
				$$\det(\lambda I-A)=\lambda^{n-v(\bar A)}\phi_A(\lambda),$$where $\phi_A(\lambda)=\lambda^{v(\bar A)}+a_1\lambda^{v(\bar A)-1}+\cdots+ a_{v(\bar A)}$. It has been proven in \citet[Lem 2]{hosoe1979irreducibility} that for almost all realizations $A$ of $\bar A$, all $v(\bar A)$ roots of $\phi_A(\lambda)$ are mutually distinct and nonzero (see Lemma \ref{hosoe_irreducibility}). Consequently, to make $A$ diagonalizable, it suffices to let the algebraic multiplicity of the zero eigenvalue of $A$, which is $n-v(\bar A)$, equal its geometric multiplicity, which is $n- {\rm grank}\, \bar A$, for almost all realizations $A$ of $\bar A$. Therefore, condition (b) is sufficient for the generic diagonalizability of $\bar A$. On the other hand, if condition (b) is not satisfied, then the zero eigenvalue has its geometric multiplicity not equaling its algebraic multiplicity for almost all realizations $A$ of $\bar A$, leading to non-diagonalizability of~$A$.
				
				(b)$\Leftrightarrow$(c): If (c) holds, then ${\rm grank}\,\bar A$ equals exactly the number of vertices that are covered by the union of disjoint cycles.  By Lemma \ref{matching-decomposition}, no decomposition of ${\mathcal G}(\bar A)$ into disjoint cycles and paths can contain disjoint cycles that can cover more vertices than ${\rm grank}\,\bar A$. As a result, ${\rm grank}\,\bar A=v(\bar A)$, which is (b). On the other hand, if (b) holds, let ${\mathcal C}$ be a union of disjoint cycles that covers the maximum number of vertices. Then, ${\mathcal C}$ corresponds to a matching ${\mathcal M}$ of ${\mathcal B}(\bar A)$ and no other edge from ${\mathcal B}(\bar A)$ can be added to ${\mathcal M}$ to form a larger matching as $|{\mathcal M}|=v(\bar A)={\rm grank}\, \bar A$, which is exactly (c).
			\end{proof}

			The following corollary is immediate from the proof of Theorem \ref{generic-diagonal-theorem}.
			\begin{corollary} \label{generic-diagonal}
				Given a structured matrix $\bar A\in \{0,*\}^{n\times n}$, either almost all realizations of $\bar A$ are diagonalizable, or almost all realizations of $\bar A$ are non-diagonalizable.
			\end{corollary}

			In view of Corollary \ref{generic-diagonal}, if $\bar A$ is not generically diagonalizable, we call it generically non-diagonalizable, reflecting the fact that almost all its realizations are non-diagonalizable. With some abuse of terminology, if $\bar A$ is generically diagonalizable, we say the digraph ${\mathcal G}(\bar A)$ is {\emph{structurally diagonalizable}}.  {Below, we present some easily verified classes of structurally diagonalizable graphs. It should be noted that the class of structurally diagonalizable graphs is much broader than the two cases mentioned in the following Corollaries \ref{corollay-acyclic} and \ref{corollay-symmetry}. For example, Figs. \ref{exam-diagonalizability-1}(a) and (b) present two structurally diagonalizable digraphs not falling into these cases. Another example is when $\bar{A}$ has full generic rank, in which case ${\mathcal G}(\bar{A})$ is called structurally cyclic in the literature \citep{moothedath2019sparsest}. Characterizing the distribution of structurally diagonalizable graphs within the set of all graphs is a subject for our future research. } %It is obvious that when each vertex of ${\mathcal G}(\bar A)$ has a self-loop, then $\bar A$ is generically diagonalizable. %An additional example is when $\bar{A}$ has full generic rank, in which case the corresponding ${\mathcal G}(\bar{A})$ is called structurally cyclic in the literature \cite{moothedath2019sparsest}.
			
						\begin{figure}[h]
				\centering
				% Requires \usepackage{graphicx}
				\includegraphics[width=3.00in]{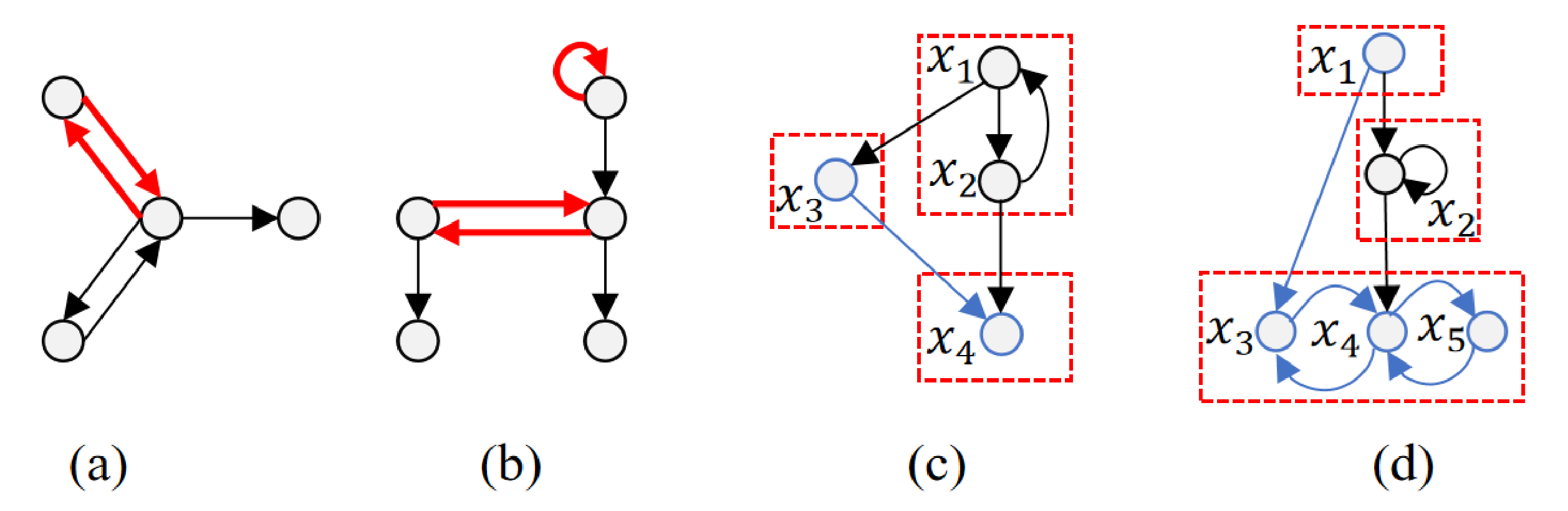}\\
			\caption{{ (a) and (b): Examples of structurally diagonalizable graphs. Bold edges correspond to a maximum matching ${\mathcal M}$, associated with which ${\mathcal G}({\mathcal M})=(X,{\mathcal M})$ is a union of disjoint cycles and isolated vertices.} (c) and (d): Examples of structurally non-diagonalizable graphs. Each SCC is in a box, while the subgraphs in blue are induced by a specific set of SCCs. In (c), the subgraph induced by $\{x_3,x_4\}$ is acyclic, and thus is structurally non-diagonalizable (Corollary \ref{corollay-acyclic}). In (d), the subgraph induced by $\{x_1,x_3,x_4,x_5\}$ is structurally non-diagonalizable as any maximum matching of it cannot correspond to disjoint cycles and isolated vertices.} \label{exam-diagonalizability-1}
			\end{figure}

			\begin{corollary}\label{corollay-acyclic} If each vertex of ${\mathcal G}(\bar A)$ has a self-loop, then $\bar A$ is generically diagonalizable.
				If ${\mathcal G}(\bar A)$ is acyclic (i.e., without any cycles, including self-loops) and $\bar A\ne 0$, then $\bar A$ is generically non-diagonalizable.
			\end{corollary}
			
			\begin{proof} The first statement is immediate from Theorem \ref{generic-diagonal-theorem} (c).
				If there is no cycle in ${\mathcal G}(\bar A)$ and $\bar A\ne 0$, then every maximum matching ${\mathcal M}$ of ${\mathcal B}(\bar A)$ corresponds to ${\mathcal G}({\mathcal M})$ with at least one path with nonzero length. According to  Theorem \ref{generic-diagonal-theorem} (c), $\bar A$ is generically non-diagonalizable.
			\end{proof}
			
			\begin{corollary}\label{corollay-symmetry}
				If $\bar A$ is structurally symmetric, possibly with nonzero diagonal entries (i.e., $\bar A_{ij}\ne 0$ implies $\bar A_{ji}\ne 0$, $\forall i,j$; equivalently, every edge of ${\mathcal G}(\bar A)$ is bidirectional except for the self-loops, or simply put, ${\mathcal G}(\bar A)$ is undirected)\footnote{This does not mean all realizations of $\bar A$ are symmetric. Instead, the symmetric realizations correspond to a set of zero measure in ${\mathbb R}^{n_{\bar A}}$.}, then $\bar A$ is generically diagonalizable.
			\end{corollary}
			\begin{proof}
				Let ${\mathcal M}$ be a maximum matching of ${\mathcal B}(\bar A)$. We first show that ${\mathcal G}({\mathcal M})=(X,{\mathcal M})$ is a union of disjoint cycles and paths, where each path has an even length. Suppose, for the sake of contradiction, there is a path $P=(x_{i_1},x_{i_2},...,x_{i_k})$ with an odd length in ${\mathcal G}({\mathcal M})$, i.e., $k\ge 2$ is even. By the symmetry of $\bar A$, there exist $k/2$ cycles in ${\mathcal G}(\bar A)$, consisting of edges $\{(x_{i_1},x_{i_2}), (x_{i_2},x_{i_1})\}$, $\cdots$, $\{(x_{i_{k-1}},x_{i_k}),(x_{i_k},x_{i_{k-1}})\}$. The {$k/2$} cycles contain $k$ edges in total, and are disjoint with the remaining cycles and paths of ${\mathcal G}({\mathcal M})$ (except for $P$). This corresponds to a matching of size $|{\mathcal M}|+1$ in ${\mathcal B}(\bar A)$, contradicting the assumption that ${\mathcal M}$ is a maximum matching. For each path with an even length in $(X,{\mathcal M})$, say $P'=(x_{i_1},x_{i_2},...,x_{i_k})$, where $k\ge 3$ is odd, due to the symmetry of $\bar A$, there are {$(k-1)/2$} cycles in ${\mathcal G}(\bar A)$, consisting of edges $\{(x_{i_2},x_{i_3}), (x_{i_3},x_{i_2})\}$, $\cdots$, $\{(x_{i_{k-1}},x_{i_k}),(x_{i_k},x_{i_{k-1}})\}$. In other words, $P'$ can be covered by a union of disjoint cycles and isolated vertices, and this union contains the same number $k-1$ of edges as $P'$. It is then easy to see that there is a maximum matching ${\mathcal M}'$ of ${\mathcal B}(\bar A)$ such that $(X,{\mathcal M}')$ is a union of disjoint cycles and isolated vertices. The required result then follows from Theorem \ref{generic-diagonal-theorem}.
			\end{proof}
			
			\begin{proposition}[Generic diagonalizability \& SCC] \label{SCC-relation}
				A structured matrix $\bar A\in \{0,*\}^{n\times n}$ is generically diagonalizable, if and only if every subgraph of ${\mathcal G}(\bar A)$ induced by the union of vertices of each subset of SCCs of ${\mathcal G}(\bar A)$ (i.e., the subgraph induced by $\bigcup_{i\in {\mathcal S}}X_i$ for each subset ${\mathcal S}\subseteq \{1,...,r\}$, where $r$ is the number of SCCs of ${\mathcal G}(\bar A)$, and $X_i$ is the vertex set of the $i$th SCC) is structurally diagonalizable.
			\end{proposition}
			
			\begin{proof} { Notice that the complete set ${\mathcal S}=\{1,...,r\}$ corresponds to exactly ${\mathcal G}(\bar A)$. The sufficiency follows immediately.} For the necessity, since ${\bar A}$ is generically diagonalizable, by (c) of Theorem \ref{generic-diagonal-theorem}, there is a maximum matching ${\mathcal M}$ of ${\mathcal B}(\bar A)$ such that ${\mathcal G}({\mathcal M})$ is the union of disjoint cycles and isolated vertices. By the definition of SCC, all edges of a cycle of ${\mathcal G}({\mathcal M})$ must belong to the same SCC. As a result, given any ${\mathcal S}\subseteq \{1,...,r\}$, the subgraph of ${\mathcal G}({\mathcal M})$ induced by the vertices $X_{\mathcal S}\doteq \bigcup_{i\in {\mathcal S}}X_i$, denoted by ${\mathcal G}({\mathcal M},{\mathcal S})$, is the union of disjoint cycles and isolated vertices, too. Denote by ${\mathcal G}({ \bar A},{\mathcal S})$ the subgraph of ${\mathcal G}(\bar A)$ induced by $X_{\mathcal S}$, and its associated bipartite graph is given by ${\mathcal B}({\bar A},{\mathcal S})$. It turns out that the set of edges of ${\mathcal G}({\mathcal M},{\mathcal S})$ corresponds to a maximum matching ${\mathcal M}_{\mathcal S}$ of ${\mathcal B}({\bar A},{\mathcal S})$. Indeed,  if there is a matching ${\mathcal M}'$ of ${\mathcal B}({\bar A},{\mathcal S})$ whose size is bigger than $|{\mathcal M}_{\mathcal S}|$, then ${\mathcal M}'\cup {\mathcal M}_{{\mathcal S}^c}$ is a matching of ${\mathcal B}(\bar A)$ with size bigger than $|{\mathcal M}|$, where ${\mathcal M}_{{\mathcal S}^c}\doteq {\mathcal M}\backslash {\mathcal M}_{\mathcal S}$, noting that the end vertices of ${\mathcal M}_{{\mathcal S}^c}$ belong to $X\backslash X_{\mathcal S}$. This  contradicts the fact that ${\mathcal M}$ is a maximum matching of ${\mathcal B}(\bar A)$. Since ${\mathcal G}({\mathcal M},{\mathcal S})$ consists of disjoint cycles and isolated vertices, by Theorem \ref{generic-diagonal-theorem} (c), ${\mathcal G}(\bar A, {\mathcal S})$ is structurally diagonalizable.
			\end{proof}
			
			%	\begin{remark}
				It is known that if ${\mathcal G}(\bar A)$ has $r$ SCCs, then there is a permutation matrix $P$ such that $P\bar A P^{-1}$ has a block-triangular form, where the $i$th diagonal block corresponds to the vertex set $X_i$ of the $i$th SCC, $i=1,...,r$ \citep{Murota_Book}. From this view, when $|{\mathcal S}|=1$, Proposition \ref{SCC-relation} coincides with the property that for a block-triangular numerical matrix to be diagonalizable, each of its diagonal blocks should be diagonalizable (this can be proved by contradiction).  However, when $|{\mathcal S}|>1$, a similar property does not necessarily hold for a numerical matrix (i.e., the property that the submatrix with rows and columns corresponding to two or more diagonal blocks should be diagonalizable). For instance, consider the matrix
				$A={\tiny{\left[\begin{array}{cc|cc|cc}
							0 & 0 & 0 & 0 & 0 & 0 \\
							0 & 0 & 0 & 0 & 0 & 0 \\
							\hline
							1 & 1 & 1 & 1 & 0 & 0 \\
							1 & 1 & 1 & 1 & 0 & 0 \\
							\hline
							1 & 1 & 1 & 1 & 0 & 0 \\
							1 & 1 & 1 & 1 & 0 & 0
						\end{array}\right]}}.$
				It turns out that $A$ is diagonalizable. However, the sub-matrix consisting of the $1st$ and $3rd$ row blocks and column blocks of $A$ is non-diagonalizable. Nevertheless, Proposition \ref{SCC-relation} implies that this property holds generically for numerical matrices.	 This proposition  provides a way to verify the generic non-diagonalizability of large-dimensional structured matrices using {\emph{local information}}, i.e., from a subset of SCCs of the associated graphs. In case ${\mathcal G}(\bar A, {\mathcal S})$ (the subgraph of ${\mathcal G}(\bar A)$ induced by $\bigcup_{i\in {\mathcal S}}X_i$) is structurally non-diagonalizable for some ${\mathcal S}\subseteq \{1,...,r\}$, it follows that $\bar A$ is so (see Figs. \ref{exam-diagonalizability-1}(c) and (d) for illustration). This proposition will play a role in proving a criterion  of SOC (Corollary \ref{corollay-diagonalizable-solvable}) in Section \ref{soc-section}.		
				%	\end{remark}
			
			% \footnote{To show this, consider $A={\tiny{\left[\begin{array}{cc|cc|cc}
							% 				0 & 0 & 0 & 0 & 0 & 0 \\
							% 				0 & 0 & 0 & 0 & 0 & 0 \\
							% 				\hline
							% 				1 & 1 & 1 & 1 & 0 & 0 \\
							% 				1 & 1 & 1 & 1 & 0 & 0 \\
							% 				\hline
							% 				1 & 1 & 1 & 1 & 0 & 0 \\
							% 				1 & 1 & 1 & 1 & 0 & 0
							% 			\end{array}\right]}}$. It turns out that $A$ is diagonalizable. However, the sub-matrix consisting of the $1st$ and $3rd$ rows and column blocks of $A$ is non-diagonalizable.}
			
			%	\begin{remark}
				
				%	\end{remark}
			
%			
%			\begin{figure}
%				\centering
%				% Requires \usepackage{graphicx}
%				\includegraphics[width=1.7in]{non-diagonal.eps}\\
%				\caption{Examples of structurally non-diagonalizable graphs. In (a) and (b), each SCC is in a box, while the subgraphs in blue are induced by a specific set of SCCs. In (a), the subgraph induced by $\{x_3,x_4\}$ is acyclic, and thus is structurally non-diagonalizable (Corollary \ref{corollay-acyclic}). In (b), the subgraph induced by $\{x_1,x_3,x_4,x_5\}$ is structurally non-diagonalizable as any maximum matching of it cannot correspond to disjoint cycles and isolated vertices. } \label{scc-diagonalizability}
%			\end{figure}

			%It is important to note that condition (c) in Theorem \ref{generic-diagonal-theorem} holds only for some maximum matching of ${\mathcal B}(\bar A)$, rather than for every maximum matching.
			
			In what follows, we provide a weighted maximum matching based algorithm to verify generic diagonalizability, which avoids enumerating all maximum matchings of ${\mathcal B}(\bar A)$ in verifying condition (c) of Theorem \ref{generic-diagonal-theorem}.
			
			\begin{proposition} \label{check-diagonalizability}
				%	There is a polynomial-time algorithm to verify whether a given structured matrix $\bar A\in \{0,*\}^{n\times n}$ is generically diagonalizable in time $O(n^3)$.
				Given a structured matrix $\bar A\in \{0,*\}^{n\times n}$, the problem of verifying whether $\bar A$ is generically diagonalizable can be reduced to a weighted maximum matching problem, and hence is solvable in $O(n^3)$ time.
			\end{proposition}
			
			\begin{proof}
				Define an auxiliary bipartite graph $\bar {\mathcal B}(\bar A)=(X,X,E_{XX}\cup E_{\lambda})$ associated with $\bar A$, where  $E_{\lambda}\!\!=\!\!\{(x_i,x_i)\!\!: \bar A_{ii}=0\}$. Assign the cost $c(e): E_{XX}\cup E_{\lambda}\to {\mathbb N}$ as follows:
				\begin{align*}
					\begin{split}	
						c(e)= \left\{			
						\begin{array}{cc}
							1,                    & {\rm if} \ e\in E_{\lambda}\\
							0,                     & {\rm if} \ e\in E_{XX}
						\end{array}
						\right.
					\end{split}
				\end{align*}
				We will show that $\bar A$ is generically diagonalizable, if and only if the minimum weight of a maximum matching of ${\bar {\mathcal B}}(\bar A)$, given by ${\rm MWMM}(\bar A)$, equals $n-{\rm grank}\,\bar A$. By condition (b) of Theorem \ref{generic-diagonal-theorem}, it suffices to prove that $v(\bar A)=n-{\rm MWMM}(\bar A)$. Observe that the size of a maximum matching of $\bar {\mathcal B}(\bar A)$ is $n$ since $\{(x_i,x_i): i=1,...,n\}$ is such a maximum matching. Notice that the disjoint cycles covering  $v(\bar A)$ vertices in ${\mathcal G}(\bar A)$ together with $n- v(\bar A)$ edges in $E_{\lambda}$ form a maximum matching of ${\bar {\mathcal B}}(\bar A)$ with weight $n-v(\bar A)$. By the assignment of edge costs, this means ${\rm MWMM}(\bar A)\le n- v(\bar A)$. Now assume ${\rm MWMM}(\bar A)< n- v(\bar A)$. Let ${\mathcal M}$ be a maximum matching of ${\bar {\mathcal B}}(\bar A)$ with ${\rm MWMM}(\bar A)$ edges belonging to $E_{\lambda}$, and denote these edges by ${\mathcal M}'\subseteq E_{\lambda}$. Let $X_1\subseteq X$ be the set of vertices left matched associated with ${\mathcal M}'$, and let $X_2\doteq X\backslash X_1$. Further, let ${\mathcal G}(\bar A, X_2)$ be the subgraph of ${\mathcal G}(\bar A)$ induced by $X_2$, and ${\mathcal B}(\bar A, X_2)$ be the biparatite graph associated with ${\mathcal G}(\bar A, X_2)$.  Observe that all edges in ${\mathcal M}\backslash {\mathcal M}'$ have end vertices in $X_2$. Consequently, ${\mathcal M}\backslash {\mathcal M}'$ is a perfect matching of ${\mathcal B}(\bar A, X_2)$, which corresponds to a union of disjoint cycles that spans ${\mathcal G}(\bar A, X_2)$ according to Lemma \ref{matching-decomposition}. By the definition of $v(\bar A)$, this leads to $v(\bar A)\ge |X_2|=n-|X_1|$, causing a contradiction to the assumption $v(\bar A)<n-|X_1|$. Therefore,  ${\rm MWMM}(\bar A)= n- v(\bar A)$. The proof is then accomplished from the fact that the weighted maximum matching problem on a bipartite graph with $n$ vertices can be solved in $O(n^3)$ time~\citep{Ahuja1993NetworkFT}.
			\end{proof}

			\section{SFO criteria for generically diagonalizable systems} \label{sec-sfo-diagonal}

		In this section, we present criteria for the SFO of generically diagonalizable systems.
		
		Some notations and concepts are introduced first. A {\emph{dilation}} in ${\mathcal G}(\bar A, \bar C)$ is a set of state vertices $X_s\subseteq X$ satisfying $|{\mathcal N}^{\rm o}(X_s)|<|X_s|$, with ${\mathcal N}^{\rm o}(X_s)\doteq\{v\in X\cup Y: (x, v)\in E_{XX}\cup E_{XY}, x\in X_s\}$ being the set of out-neighbors of $X_s$. A {\emph{minimal dilation}} is a dilation such that any of its proper subsets is not a dilation. Given $\bar F\in \{0,*\}^{r\times n}$, let $X_F=\{x_i\in X: \bar F(:,i)\ne 0\}$ be the set of state vertices corresponding to nonzero columns of $\bar F$, i.e., states that contribute directly to $z(t)$, which are called {\emph{functional states}}. Let $\bar e_i\in \{0,*\}^{1\times n}$ be such that the $i$th entry of $\bar e_{i}$ is $*$ and the rest are zero. Define $\bar I_{X_F}\doteq {\bf col}\{\bar e_i: x_i\in X_F\}$, and $\bar I_n\doteq{\bf col}\{\bar e_i: x_i\in X\}$.
		
		Given a numerically specific triple $(A,C,F)$,  let $\lambda$  be an eigenvalue of $A$.  If the algebraic multiplicity of $\lambda$ equals its geometric multiplicity, then $\lambda$ is said to be {\emph{modal functionally observable}} if condition (\ref{rank-criterion-observ}) holds for this particular $\lambda$ \citep{zhang2023functional}. If the algebraic multiplicity of $\lambda$ is greater than its geometric multiplicity, the definition of modal functional observability involves the Jordan normal form of $A$ and the associated observability matrices; see \cite{zhang2023functional} for details. It has been shown in \cite{zhang2023functional} that $(A,C,F)$ is functionally observable, if and only if each eigenvalue of $A$ is modal functionally observable. Below, we show that the output-reachability of $X_F$ in ${\mathcal G}(\bar A, \bar C)$ implies that every nonzero eigenvalue of $A$ is generically modal functionally observable, generalizing \citet[Theo 2]{hosoe1979irreducibility}.

		\begin{proposition} \label{nonzero-mode-sfo}
			If every $x\in X_F$ is output-reachable, then for almost all realizations $(A,C,F)$ of $(\bar A, \bar C, \bar F)$, every nonzero eigenvalue of $A$ is modal functionally observable, {that is, there is a proper variety ${\mathbb V}\subseteq {\mathbb R}^{n_{\bar A}}\times {\mathbb R}^{n_{\bar C}}\times {\mathbb R}^{n_{\bar F}}$, such that for any $(A,C,F)\in {\mathbb R}^{n_{\bar A}}\times {\mathbb R}^{n_{\bar C}}\times {\mathbb R}^{n_{\bar F}}\backslash {\mathbb V}$, ${\rm rank}[A-\lambda I; C; F]={\rm rank}[A-\lambda I; C]$, where $\lambda$ is any nonzero eigenvalue of $A$.}
		\end{proposition}
		
		\begin{proof}
			Without loss of generality, assume that vertices $X_1\doteq \{x_1,...,x_{n_1}\}$ are output-reachable while $X_2\doteq X\backslash X_1= \{x_{n_1+1},...,x_n\}$ are not, where $n_1=|X_1|$. If every $x\in X_F$ is output-reachable, i.e., $X_F\subseteq X_1$, then for all realizations $(A,C,F)$ of $(\bar A, \bar C, \bar F)$, $[A-\lambda I; C; F]$ has the following form
			{\small	$$\left[\begin{array}{cc}
					A_{11}-\lambda I & 0 \\
					A_{21} & A_{22}-\lambda I \\
					C_1 & 0 \\
					F_1 & 0
				\end{array}\right],$$}where $A_{11}\in {\mathbb R}^{n_1\times n_1}$, $A_{22}\in {\mathbb R}^{(n-n_1)\times (n-n_1)}$, and $[C_1,0]$ and $[F_1, 0]$ are partitioned in accordance with $A$.
			From Lemma \ref{hosoe_irreducibility}, there is a proper variety ${\mathbb V}\subseteq {\mathbb R}^{n_{\bar A}}\times {\mathbb R}^{n_{\bar C}}\times {\mathbb R}^{n_{\bar F}}$ such that for every $(A,C,F)\in {\mathbb R}^{n_{\bar A}}\times {\mathbb R}^{n_{\bar C}}\times {\mathbb R}^{n_{\bar F}}\backslash {\mathbb V}$, all nonzero eigenvalues of $A$ (if any) are simple, and
			${\rm rank}[A_{11}-\lambda I; C_1]=n_1$ for each nonzero eigenvalue $\lambda$ of $A_{11}$. This leads to that
			${\rm rank}[A_{11}-\lambda I; C_1]=n_1, \forall \lambda\ne 0$. As a result, for all $\lambda \in {\mathbb C}\backslash \{0\}$,
			{\small	$${\rm rank}\left[\begin{array}{cc}
					A_{11}-\lambda I & 0 \\
					A_{12} & A_{22}-\lambda I \\
					C_1 & 0 \\
					F_1 & 0
				\end{array}\right]\!=\!{\rm rank}\left[\begin{array}{cc}
					A_{11}-\lambda I & 0 \\
					A_{12} & A_{22}-\lambda I\\
					C_1 & 0 \\
					0 & 0
				\end{array}\right],$$}which is due to the fact that ${\mathcal R}(F_1)\subseteq {\mathcal R}([A_{11}-\lambda I; C_1])$, with ${\mathcal R}(M)$ denoting the space spanned by the rows of a matrix $M$. The above relation is exactly the definition of modal functional observability for a simple eigenvalue $\lambda$ of $A$, leading to the proposed statement.
		\end{proof}

		\begin{theorem} \label{main-theorem}
			Given a triple $(\bar A, \bar C, \bar F)$, suppose  $\bar A$ is generically diagonalizable. Then, the following statements are equivalent:
			\begin{itemize}
				\item[(a)] $(\bar A, \bar C, \bar F)$ is SFO.
				\item[(b)] Every $x\in X_F$ is output-reachable (b1), and \\ ${\rm grank}\,[\bar A;\bar C]={\rm grank}\,[\bar A;\bar C; \bar F]$ (b2).
				\item[(c)] Every $x\in X_F$ is output-reachable (c1), and \\  ${\rm grank}\,[\bar A;\bar C]={\rm grank}\,[\bar A;\bar C; \bar e_i]$ for each $x_i\in X_F$ (c2).
				\item[(d)] Every $x\in X_F$ is output-reachable (d1), and no $x\in X_F$ is contained in a minimal dilation of ${\mathcal G}(\bar A, \bar C)$ (d2).
			\end{itemize}
		\end{theorem}

		\begin{proof}
			(b)$\Rightarrow$ (a): From Proposition \ref{nonzero-mode-sfo}, condition (b1) implies that there is a proper variety ${\mathbb V}_1$ such that for all $(A,B,C)\in {\mathbb R}^{n_{\bar A}}\times {\mathbb R}^{n_{\bar C}}\times {\mathbb R}^{n_{\bar F}}\backslash {\mathbb V}_1$, ${\rm rank}[A-\lambda I; C; F]={\rm rank}[A-\lambda I; C]$, $\forall \lambda\in {\mathbb C}\backslash \{0\}$. Condition (b2) implies that there is a proper variety ${\mathbb V}_2$ such that for all $(A,B,C)\in {\mathbb R}^{n_{\bar A}}\times {\mathbb R}^{n_{\bar C}}\times {\mathbb R}^{n_{\bar F}}\backslash {\mathbb V}_2$, ${\rm rank}[A;C]={\rm rank}[A;C;F]$. Since $\bar A$ is generically diagonalizable, by definition, there exists a proper variety ${\mathbb V}_3$ such that for all $(A,B,C)\in {\mathbb R}^{n_{\bar A}}\times {\mathbb R}^{n_{\bar C}}\times {\mathbb R}^{n_{\bar F}}\backslash {\mathbb V}_3$, $A$ is diagonalizable. Those conditions together yield that for the proper variety ${\mathbb V}_4={\mathbb V}_1\cup {\mathbb V}_2\cup {\mathbb V}_3$,
			all realizations $(A,C,F)\in {\mathbb R}^{n_{\bar A}}\times {\mathbb R}^{n_{\bar C}}\times {\mathbb R}^{n_{\bar F}}\backslash {\mathbb V}_4$ satisfy that $A$ is diagonalizable, and  $${\rm rank}[A-\lambda I; C; F]\!=\!{\rm rank}[A-\lambda I; C], \forall \lambda\in {\mathbb C},$$which indicates that $(A,C,F)$ is functionally observable from Lemma \ref{functional-theorem-diagonal}. 	By definition, $(\bar A, \bar C, \bar F)$ is SFO.
			%				$$\begin{array}{c}
				%					A \ {\rm diagonalizable,} \ {\rm and} \\
				%					{\rm rank}[A-\lambda I; C; F]\!=\!{\rm rank}[A-\lambda I; C], \forall \lambda\in {\mathbb C},
				%				\end{array}$$

			(a)$\Rightarrow$(b): If (b1) is not satisfied, without loss of generality, assume that $X_1=\{x_1,...,x_{n_1}\}$ is the set of output-reachable vertices, and $X_2=X\backslash X_1=\{x_{n_1+1},...,x_n\}$. It follows that $X_F\cap X_2\ne \emptyset$. Accordingly, partition $(\bar A,\bar C,\bar F)$ as
			{\small	$$\left[\begin{array}{c}
					\bar A \\
					\bar C \\
					\bar F
				\end{array}\right]=\left[\begin{array}{cc}
					\bar A_{11} & 0 \\
					\bar A_{12} & \bar A_{22} \\
					\hline
					\bar	C_1 & 0 \\
					\hline
					\bar 	F_1 & \bar  F_2
				\end{array}\right],$$}with $\bar A_{11}\in\{0,*\}^{n_1\times n_1}$, $\bar A_{22}\in \{0,*\}^{(n-n_1)\times (n-n_1)}$, and $\bar F_2\ne 0$. Then, ${\rm grank}\,O(\bar A,\bar C)={\rm grank}\,O(\bar A_{11}, \bar C_1)$, ${\rm grank}\,O(\bar A,[\bar C;\bar F])\ge {\rm grank}\,O(\bar A_{11}, \bar C_1)+ {\rm grank}\, O(\bar A_{22}, \bar F_2)$. As $\bar F_2\ne 0$, ${\rm grank}\, O(\bar A_{22}, \bar F_2)\ge 1$, yielding ${\rm grank}\,O(\bar A,[\bar C;\bar F])>{\rm grank}\,O(\bar A,\bar C).$ By Lemma \ref{characterization-pro}, $(\bar A, \bar C, \bar F)$ is not SFO, which demonstrates the necessity of (b1) for (a). The necessity of (b2) for (a) is obvious since condition (\ref{rank-criterion-observ}) should hold when $\lambda =0$.
			
			To show (b)$\Leftrightarrow$(c), it suffices to show (b2)$\Leftrightarrow$(c2). The direction (b2)$\Rightarrow$(c2) is obvious by noting that $\forall x_i\in X_F$,
			${\rm grank}\,[\bar A; \bar C; \bar F]\ge {\rm grank}\,[\bar A; \bar C; \bar e_i]\ge {\rm grank}\,[\bar A; \bar C].$
			
			(c2)$\Rightarrow$(b2): By the property of rank function\footnote{The generic rank function has the submodularity property \citep{Murota_Book}, that is, given $\bar M\in \{0,*\}^{m\times n}$, for any $S_1,S_2\subseteq [m]$, ${\rm grank}\,\bar M(S_1,:)+{\rm grank}\,\bar M(S_2,:)\ge {\rm grank}\,\bar M(S_1\cup S_2,:)+{\rm grank}\,\bar M(S_1\cap S_2,:)$.}, if ${\rm grank}\,[\bar A; \bar C]$ $={\rm grank}\,[\bar A;\bar C;\bar e_i]={\rm grank}\,[\bar A;\bar C;\bar e_j]$ whenever $i\ne j$, then ${\rm grank}\,[\bar A; \bar C]={\rm grank}\,[\bar A;\bar C;\bar e_i;\bar e_j]$. Repeat this procedure, and we get ${\rm grank}\,[\bar A; \bar C]={\rm grank}\,[\bar A;\bar C;\bar I_{X_F}]$. Let $X_F^c=X\backslash X_F$. Observe that
			${\rm grank}\,[\bar A; \bar C;\bar I_{X_F}]={\rm grank}\,[\bar A; \bar C](:,X_F^c)+|X_F|$, and
			${\rm grank}\,[\bar A; \bar C] \!\le\! {\rm grank}\,[\bar A; \bar C; \bar F]\!\le \!{\rm grank}\,[\bar A; \bar C](:,X_F^c)+|X_F|$.
			It follows that ${\rm grank}\,[\bar A; \bar C]={\rm grank}\,[\bar A; \bar C; \bar F]$, yielding (c2).
			
			To show (c)$\Leftrightarrow$(d), it suffices to show (c2)$\Leftrightarrow$(d2), which has been proved in \citet[Lems 2, 5]{montanari2022functional}.
		\end{proof}

		%	It is worth mentioning that condition (d) was posited as both necessary and sufficient for the SFO of a general structured system in \cite{montanari2022functional}. However, this assertion is not accurate due to the adoption of the PBH-like necessary condition (condition (\ref{rank-criterion-observ})) as a sufficient condition for functional observability in the general case, {which was originated in \cite{moreno2001quasi,jennings2011existence} and not corrected until \cite{asymptotic2022Mohamed,functional2022Mohamed,zhang2023functional}}.
%it appears there might have been some misunderstanding in treating the PBH-like necessary condition (condition (\ref{rank-criterion-observ})) as a sufficient condition for functional observability in the general case. This perspective, which originally appeared in \cite{moreno2001quasi,jennings2011existence}, was not clarified until recently \cite{asymptotic2022Mohamed,functional2022Mohamed,zhang2023functional}.
	\begin{remark}\label{comparison-existing}
It is worth mentioning that condition (d) was posited as both necessary and sufficient for the SFO of a general structured system in \cite{montanari2022functional}. { However, it appears there might have been some misunderstanding in treating the PBH-like necessary condition (condition (\ref{rank-criterion-observ})) as a sufficient condition for functional observability in the general case. This perspective, which originally appeared in \cite{moreno2001quasi} and \cite{jennings2011existence}, was not clarified until recently \citep{asymptotic2022Mohamed,functional2022Mohamed,zhang2023functional}.} Theorem \ref{main-theorem} suggests that the conditions obtained in \cite{montanari2022functional} are applicable to the class of generically diagonalizable systems. It is essential to underscore several key points:
			%	Theorem \ref{main-theorem} suggests that, within the class of generically diagonalizable systems, the conditions obtained in \cite{montanari2022functional} are correct. Nevertheless, it is essential to underscore several key points:
			\begin{itemize}
				\item Firstly, \cite{montanari2022functional} derived condition (d) under a different definition of SFO compared to ours.  In their definition, a structured system is considered SFO if there exists a realization that is functionally observable, rather than in the generic case. As outlined in Definition \ref{def-structural-functional} and demonstrated in {Example \ref{example-counter}}, these two definitions are not equivalent, {even for generically diagonalizable systems (notice that $\bar A$ in Example \ref{example-counter} is generically diagonalizable)}. This variance in the definitions of SFO leads to distinct derivations of SFO criteria.
			%	\item Secondly, there is no guarantee that the graph-theoretic condition (d) is equivalent to condition (\ref{rank-criterion-observ}) in the generic sense if no constraint is imposed on $\bar A$ (see Remark 7 of \cite{zhang2023functional}). {It is worth mentioning that \citet[Theo 1]{montanari2024popov} has recently proposed additional conditions to ensure such equivalence for a numerical system. These conditions depend on the matrices $A$ and $F$ without requiring the diagonalizability of $A$.}
				% unless the generic diagonalizability condition on $\bar A$ is satisfied. Such equivalence was reported in \cite{montanari2022functional} without imposing any particular constraint on $\bar A$.

				\item {Secondly, condition (d) given in \cite{montanari2022functional} was predicated on an additional assumption -- that each row of $\bar F$ contains only one nonzero entry. This assumption is not needed here.}
				
				\item Thirdly, there is no guarantee that the {graph-theoretic} condition (d) is equivalent to condition (\ref{rank-criterion-observ}) in the generic sense if no constraint is imposed on $\bar A$ (see Remark 7 of \cite{zhang2023functional}). Building on Proposition \ref{nonzero-mode-sfo}, it may be possible to broaden Theorem \ref{main-theorem} to encompass more general systems. However, the outcomes would be contingent on $\bar A, \bar C$, and $\bar F$, rather than solely on $\bar A$. {In particular, \citet[Theo 1]{montanari2024popov} recently proposed additional conditions to ensure the equivalence between condition (\ref{rank-criterion-observ}) of Lemma \ref{functional-theorem-diagonal} and condition (ii) of Lemma \ref{characterization-pro} for a numerical system, depending on the matrices $A$ and $F$ without requiring the diagonalizability of $A$. These results may be promising in deriving graph-theoretic conditions for SFO like Theorem~\ref{main-theorem}.}
				
				 %{see \citet[Theo 1]{montanari2024popov} for some discussions for numerical systems}.
				
				%%				\item Firstly, condition (d) is not equivalent to condition (\ref{rank-criterion-observ}) in the generic sense (see Remark 7 of \cite{zhang2023functional}) unless the generic diagonalizability condition on $\bar A$ is satisfied. Such equivalence was reported in \cite{montanari2022functional} without imposing any particular constraint on $\bar A$.
				%				\item Secondly, \cite{montanari2022functional} derived condition (d) under a different definition of SFO compared to ours.  In their definition, a structured system is considered SFO if there exists a realization that is functionally observable, rather than in the general case (as outlined in Definition \ref{def-structural-functional} and discussed in Remark \ref{example-counter}). This variance in the definitions of SFO leads to distinct derivations of SFO criteria.
				%				\item Lastly, condition (d) given in \cite{montanari2022functional} was predicated on two additional assumptions: 1) that each row of $\bar F$ contains only one nonzero entry, and 2) that $[\bar C;\bar F]$ possesses full row generic rank (referred to as assumption 1) and 2)). These assumptions are not needed in our results.
			\end{itemize}
		\end{remark}

		Compared to Lemma \ref{characterization-pro} or the graph-theoretic conditions in \cite{zhang2023functional}, Theorem \ref{main-theorem} stands out for its remarkable simplicity as it does need the global information of $(\bar A, \bar C, \bar F)$ (such as the {generic dimension of unobservable subspaces, i.e., $n-{\rm grank}\,O(\bar A, \bar C)$}). This notable feature enables the derivation of a closed-form solution for the minimal sensor placement problem within the class of generically diagonalizable systems, {\emph{a problem that remains open otherwise}} (see Section \ref{sec-sensor}).

			%\begin{proof} By regarding $O(\bar A, \bar C)$ as $[\bar A;\bar C]$ in the proof of `(b)$\Leftrightarrow(c)$' of Theorem \ref{main-theorem}, we can prove ${\rm grank}\,[O(\bar A, \bar C);\bar F]={\rm grank}\,[O(\bar A, \bar C)]$, if and only if ${\rm grank}\,[O(\bar A,\bar C)]=$\\${\rm grank}\,[O(\bar A, \bar C);\bar e_i]$, $\forall x_i\in X_F$.
			%The proof then can be completed by invoking condition (i) of Lemma \ref{characterization-pro}.
			
		  	\section{Polynomial-time verifiable criteria for SOC} \label{soc-section}
		  In this section, we identify a class of systems for which the SOC can be checked in polynomial time and highlight that generically diagonalizable systems fall into it.
		
		  %	Recall $Q(A,B)=[B,AB,\cdots, A^{n-1}B]$.
		  Let $X_r$ be the set of input-reachable state vertices in ${\mathcal G}(\bar A, \bar B)$. Let $\bar A_r\in \{0,*\}^{n\times n}$ be obtained from $\bar A$ by replacing its rows and columns corresponding to $X\backslash X_r$ with zero blocks. Construct a directed graph ${\mathcal D}(\bar A_r,\bar B,\bar C)$ associated with $(\bar A_r,\bar B,\bar C)$ as follows: the vertex set is $X^2\cup U^2\cup X^1\cup Y^0$, with $X^2=\{x_1^2,\cdots, x_n^2\}$, $U^2=\{u_1,...,u_m\}$, $X^1=\{x_1^1,\cdots, x_n^1\}$, and $Y^0=\{y_1,...,y_p\}$, and the edge set is $E_{21}\cup E_{10}$, with $E_{21}=\{(x_i^2,x_j^1):[\bar A_r]_{ji}\ne 0\}\cup \{(u_i,x_j^1):\bar B_{ji}\ne 0\}$ and $E_{10}=\{(x_i^1,y_j):\bar C_{ji}\ne 0\}$. A linking of size $l$ from $X^2\cup U^2$ to $Y^0$ in ${\mathcal D}(\bar A_r,\bar B,\bar C)$ is a collection of $l$ vertex-disjoint simple paths, in which each path has its initial vertex in $X^2\cup U^2$ and terminal vertex in $Y^0$. {To illustrate these definitions, below we provide an example.}

{
\begin{example}\label{illustrate-example}
Consider a system $(\bar A, \bar B, \bar C)$ whose digraph ${\mathcal G}(\bar A, \bar B, \bar C)$ is given in Fig. \ref{soc-fig}(a). From it, we known state vertices $X_r=\{x_1,x_2,x_3,x_4\}$ are input-reachable, while $x_5$ is not. Accordingly, the corresponding ${\mathcal D}(\bar A_r,\bar B, \bar C)$ is given in Fig. \ref{soc-fig}(b), where $(x_5^2,x_4^1)$ does not exist since $x_5$ is not input-reachable. Two disjoint paths $(x_1^2,x_4^1,y_2)$ and $(x_2^2,x_3^1,y_1)$ constitute a linking of size $2$ from $X^2\cup U^2$ to~$Y^0$.
\end{example}
}
		  \begin{figure}[H]
		  	\centering
		  	% Requires \usepackage{graphicx}
		  	\includegraphics[width=2.2in]{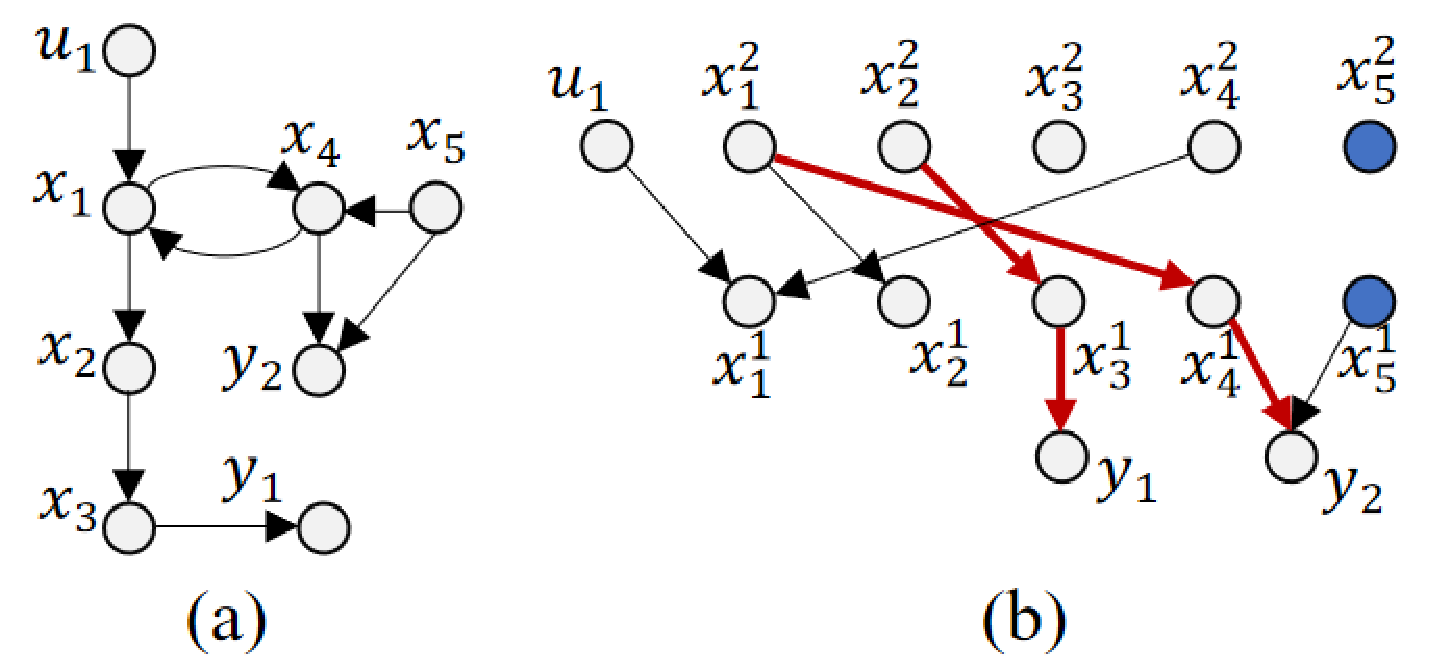}\\
		  	\caption{{${\mathcal G}(\bar A, \bar B, \bar C)$ and ${\mathcal D}(\bar A_r,\bar B,\bar C)$ in Example \ref{illustrate-example}. Vertices in blue correspond to input-unreachable vertices in ${\mathcal G}(\bar A, \bar B)$. Bold red edges constitute a linking of size $2$.}} \label{soc-fig}
		  \end{figure}
		
		  \begin{theorem} \label{solvable-index}
		  	If ${\rm grank}\, [\bar A_r, \bar B]= {\rm grank}\, Q(\bar A, \bar B)$, then $(\bar A, \bar B, \bar C)$ is SOC, if and only if there is a linking of size $p$ from $X^2\cup U^2$ to $Y^0$ in ${\mathcal D}(\bar A_r,\bar B,\bar C)$.
		  	%	$\bar C[\bar A_r, \bar B]$ has full row generic rank.
		  \end{theorem}
		  \begin{proof}
		  	From \citet[Theo. 6]{papadimitriou1984simple}, there is a linking of size $p$ from $X^2\cup U^2$ to $Y^0$ in ${\mathcal D}(\bar A_r,\bar B,\bar C)$, if and only if ${\rm grank}\, \bar C[\bar A_r,\bar B]=p$. To proceed with the proof, first consider the case where all state vertices are input-reachable, i.e., the case $\bar A_r=\bar A$. As ${\rm grank}\, [\bar A, \bar B]= {\rm grank}\, Q(\bar A, \bar B)$, there is a proper variety ${\mathbb V}\subseteq {\mathbb R}^{n_{\bar A}}\times {\mathbb R}^{n_{\bar B}}\times {\mathbb R}^{n_{\bar C}}$ such that for any $(A,B,C)\in {\mathbb R}^{n_{\bar A}}\times {\mathbb R}^{n_{\bar B}}\times {\mathbb R}^{n_{\bar C}}\backslash {\mathbb V}$, it holds that ${\rm rank}[A,B]={\rm rank} Q(A,B)$. Observe that for any $z\in {\mathbb C}^{1\times n}$ satisfying $z[A,B]=0$, we have $zA^kB=0$ for $k=0,1,\cdots, n-1$, which yields $zQ(A,B)=0$. That is, upon letting ${{\mathcal {N}}}(\cdot)$ be the left-null space of a matrix, we have
		  	$ {{\mathcal {N}}}([A,B])\subseteq {{\mathcal {N}}}(Q(A,B)).$
		  	Since ${\rm rank}[A,B]={\rm rank} Q(A,B)$, which implies ${\rm dim}{{\mathcal {N}}}([A,B])={\rm dim} {{\mathcal {N}}}(Q(A,B))$ (${\rm dim}(\cdot)$ takes the dimension of a subspace), we arrive at
		  	\begin{equation} \label{null-equalence} {{\mathcal {N}}}([A,B])={{\mathcal {N}}}(Q(A,B)).\end{equation}
		  	If $(A,B,C)$ is not { output controllable}, there exists some nonzero $z\in {\mathbb C}^{1\times p}$ such that $zCQ(A,B)=0$. That is, $zC\in {{\mathcal {N}}}(Q(A,B))$. Relation (\ref{null-equalence}) yields $zC[A,B]=0$, which implies ${\rm rank}C[A,B]<p$.
		  	Conversely, if $C[A,B]$ does not have full row rank, there exists some nonzero $z'\in {\mathbb C}^{1\times p}$ such that $z'C[A,B]=0$. This means $z'C\in {{\mathcal {N}}}([A,B])$, leading to $z'CQ(A,B)=0$ from (\ref{null-equalence}), i.e., $(A,B,C)$ is not { output controllable}.
		  	The above analysis shows that for any $(A,B,C)\in {\mathbb R}^{n_{\bar A}}\times {\mathbb R}^{n_{\bar B}}\times {\mathbb R}^{n_{\bar C}}\backslash {\mathbb V}$, $(A,B,C)$ is { output controllable}, if and only if ${\rm rank} C[A,B]=p$.
		  	As a result, if ${\rm grank}\, \bar C[\bar A,\bar B]=p$, there is a proper variety ${\mathbb V}'\subseteq {\mathbb R}^{n_{\bar A}}\times {\mathbb R}^{n_{\bar B}}\times {\mathbb R}^{n_{\bar C}}$ such that for any $(A,B,C)\in {\mathbb R}^{n_{\bar A}}\times {\mathbb R}^{n_{\bar B}}\times {\mathbb R}^{n_{\bar C}}\backslash {\mathbb V}'$, ${\rm rank}C[A,B]=p$. Moreover, for each $(A,B,C)\in {\mathbb R}^{n_{\bar A}}\times {\mathbb R}^{n_{\bar B}}\times {\mathbb R}^{n_{\bar C}}\backslash ({\mathbb V}\cup{\mathbb V}')$, $(A,B,C)$ is { output controllable}. Conversely, if ${\rm grank}\, \bar C[\bar A,\bar B]<p$, then for all $(A,B,C)\in {\mathbb R}^{n_{\bar A}}\times {\mathbb R}^{n_{\bar B}}\times {\mathbb R}^{n_{\bar C}}\backslash {\mathbb V}$, $(A,B,C)$ is not { output controllable} since ${\rm rank}C[A,B]<p$. By definition, $(\bar A,\bar B,\bar C)$ is SOC, if and only if ${\rm grank}\, \bar C[\bar A,\bar B]=p$.
		  	
		  	Now consider the case where not all state vertices are input-reachable in ${\mathcal G}(\bar A,\bar B)$. Assume $X_{r}\subseteq X$ is the set of input-reachable vertices in ${\mathcal G}(\bar A,\bar B)$. Without losing generality, suppose $[\bar A,\bar B]$ and $A_r$ have the following forms, respectively (if not, one can re-order the state vertices):
		  	\begin{equation}\label{reducible-form} [\bar A, \bar B]=\left[\begin{array}{cc|c}
		  			\bar A_{11} & \bar A_{12} & \bar B_1 \\
		  			0 &  \bar A_{22} & 0
		  		\end{array}\right],A_r=\left[\begin{array}{cc}
		  			\bar A_{11} & 0 \\
		  			0  & 0
		  		\end{array}\right],\end{equation}
		  	where $\bar A_{11}\in \{0,*\}^{|X_r|\times |X_r|}$, $\bar A_{22}\in \{0,*\}^{|X\backslash X_r|\times |X\backslash X_r|}$, and $\bar B_1\in \{0,*\}^{|X_r|\times m}$. Moreover, partition $\bar C$ as $\bar C=[\bar C_1,\bar C_2]$ with $\bar C_1\in \{0,*\}^{p\times |X_r|}$ and $\bar C_2\in \{0,*\}^{p\times |X\backslash X_r|}$. It turns out that
		  	\begin{equation}\label{partition-equality}Q(\bar A, \bar B)=\left[\begin{array}{cccc}\bar B_1 & \bar A_{11}\bar B_1 & \cdots & \bar A_{11}^{n-1}\bar B_1\\
		  		0 & 0 & \cdots & 0	
		  	\end{array}\right], \bar A_{r}=\left[ \begin{array}{cc}
		  		\bar A_{11} & 0\\
		  		0 & 0
		  	\end{array} \right].\end{equation}Therefore, ${\rm grank}\, [\bar A_r, \bar B]= {\rm grank}\, Q(\bar A, \bar B)$, if and only if ${\rm grank}\, [\bar A_{11}, \bar B_1]= {\rm grank}\, Q(\bar A_{11}, \bar B_1)$. Moreover, ${\rm grank}\,\bar CQ(\bar A, \bar B)={\rm grank}\,\bar C_1Q(\bar A_{11}, \bar B_{1})$ and ${\rm grank}\,\bar C[\bar A_r,\bar B]={\rm grank}\,\bar C_1[\bar A_{11},\bar B_{1}]$. Observe that every state vertex of $(\bar A_{11},\bar B_1)$ is input-reachable. By adopting the above SOC condition on $(\bar A_{11}, \bar B_1, \bar C_1)$, we obtain the required result.
		  \end{proof}
		
		  In light of Theorem \ref{solvable-index}, whenever ${\rm grank}\, [\bar A_r, \bar B]= {\rm grank}\, Q(\bar A, \bar B)$, we can check the SOC of $(\bar A, \bar B, \bar C)$ by verifying whether
		  a linking of size $p$ exists in ${\mathcal D}(\bar A_r, \bar B, \bar C)$. All these conditions can be verified in polynomial time. To be specific,  {${\rm grank}\, [\bar A_r, \bar B]$ is obtainable by computing the maximum size of matchings in the bipartite graph associated with $[\bar A_r, \bar B]$.} And ${\rm grank}\,Q(\bar A, \bar B)$ equals the maximum number of state vertices that can be covered by an input cactus configuration in ${\mathcal G}(\bar A, \bar B)$ \citep{hosoe1980determination}, which can be determined via weighted maximum matching algorithms \citep[Theo. 6]{murota1990note}. Moreover, the maximum size of linkings in ${\mathcal D}(\bar A_r, \bar B, \bar C)$ can be obtained via the maximum flow algorithm in $O(n^{2.5})$ time (\citet[Theo. 6]{papadimitriou1984simple}; see also Section \ref{soc-optimization}).

\begin{example}[Example \ref{illustrate-example} continued]\label{soc-example}
{Consider the system $(\bar A, \bar B, \bar C)$ given in Example \ref{illustrate-example}. Since the maximum number of input-reachable state vertices in a collection of disjoint input stems and cycles is $3$, we have ${\rm grank}\,Q(\bar A, \bar B)=3$. Moreover,  ${\rm grank}\,[\bar A_r,\bar B]=3$, leading to ${\rm grank}\,Q(\bar A, \bar B)={\rm grank}\,[\bar A_r,\bar B]$. As there is a linking of size $2$ from $X^2\cup U^2$ to $Y^0$ in ${\mathcal G}(\bar A_r,\bar B, \bar C)$ (highlighted in bold red lines in Fig. \ref{illustrate-example}(b)), Theorem \ref{solvable-index} yields that $(\bar A, \bar B, \bar C)$ is SOC.}
\end{example}

%\begin{example}[Example \ref{illustrate-example} continued]\label{soc-example}
%	Consider the system $(\bar A, \bar B, \bar C)$ given in Example \ref{illustrate-example}. {It can be verified that ${\rm grank}\,Q(\bar A, \bar B)=3={\rm grank}\,[\bar A_r,\bar B]$. As there is a linking of size $2$ from $X^2\cup U^2$ to $Y^0$ in ${\mathcal G}(\bar A_r,\bar B, \bar C)$ (highlighted in bold red lines in Fig. \ref{illustrate-example}(b)), Theorem \ref{solvable-index} yields that $(\bar A, \bar B, \bar C)$ is SOC.}

% (notice that ${\rm grank}\,[\bar A,\bar B]=4$, highlighting the necessity of using $\bar A_r$ instead of $\bar A$) For $(\bar A, \bar B, \bar C)$ with its ${\mathcal G}(\bar A, \bar B, \bar C)$ given in Fig. \ref{soc-fig}(b), it turns out that $\bar A$ is generically diagonalizable. From Corollary \ref{corollay-diagonalizable-solvable}, $(\bar A, \bar B, \bar C)$ is SOC. Moreover, for any $\bar B$ and $\bar C$, we can verify that ${\rm grank}\,Q(\bar A, \bar B)={\rm grank}\,[\bar A_r,\bar B]$, and the SOC of $(\bar A, \bar B, \bar C)$ can be checked by computing the linking size condition.
		  	%				Example 1: $\bar A$ not diagonalizable but $rank Q(A,B)=[A_r,B]$. Notice that states $\{x_3,x_4\}$ are not covered by any cactus configuration of ${\mathcal G}(\bar A, \bar B)$.
		  	%				Example 2: $\bar A$ diagonalizable.
		
{We emphasize that $\bar A$ in Example \ref{soc-example} is not generically diagonalizable and the condition ${\rm grank}\, [\bar A_r, \bar B]= {\rm grank}\, Q(\bar A, \bar B)$ depend on both $\bar A$ and $\bar B$.} In what follows, we show that for the class of generically diagonalizable systems, this condition always holds irrespective of~$\bar B$.

		  % Therefore, Theorem \ref{solvable-index} provides a scenario where SOC can be verified in polynomial time by deterministic combinatorial algorithms. %Furthermore, notice that it always holds that ${\rm grank}\, [\bar A_r, \bar B]\ge {\rm grank}\, Q(\bar A, \bar B)$.

		  \begin{corollary} \label{corollay-diagonalizable-solvable}
		  	If $\bar A$ is generically diagonalizable, then ${\rm grank}\, [\bar A_r, \bar B]= {\rm grank}\, Q(\bar A, \bar B)$ for any $\bar B\in \{0,*\}^{n\times m}$. Consequently, $(\bar A, \bar B, \bar C)$ is SOC, if and only if there is a linking of size $p$ from $X^2\cup U^2$ to $Y^0$ in ${\mathcal D}(\bar A_r,\bar B,\bar C)$. %$\bar C[\bar A_r, \bar B]$ has full row generic rank.
		  \end{corollary}
		
		  \begin{proof}
		  	We first consider the case where every state vertex in ${\mathcal G}(\bar A,\bar B)$ is input-reachable, i.e., the case $\bar A_r=\bar A$. By Lemma \ref{hosoe_irreducibility}, for a generically diagonalizable $\bar A$, there is a proper variety ${\mathbb V}$ such that for each $(A,B)\in {\mathbb R}^{n_{\bar A}}\times {\mathbb R}^{n_{\bar B}}\backslash {\mathbb V}$, it holds that (i) $A$ is diagonalizable and (ii) each nonzero eigenvalue of $A$ is simple and controllable for system $(A,B)$. We are to show that ${\rm rank} [A,B]={\rm rank} Q(A,B)$.
		  	Suppose $A$ has $q$ distinct eigenvalues $\lambda_1,...,\lambda_q$, in which $\lambda_1=0$, $\lambda_2,\cdots, \lambda_q\ne 0$. Moreover, let $z_i\in {\mathbb C}^{n_i\times n}$ consist of a set of linearly independent row vectors spanning the left eigenspace associated with $\lambda_i$, where $n_i$ is the geometric multiplicity of $\lambda_i$, and $Z$ be stacked by $z_1,...,z_q$ from up to down (if all eigenvalues of $A$ are nonzero, then $\lambda_1=0$ does not exist and $n_1=0$, under which all the subsequent statements remain valid). Then, $ZAZ^{-1}={\bf diag}\{\lambda_iI_{n_i}|_{i=1}^q\}$. According to \citet[Lem. 8]{A.Ol2014Minimal}, we have\begin{align*}
		  		{\rm rank} Q(A,B) &= \sum\nolimits_{i=1}^q {\rm rank} Q(\lambda_iI_{n_i},z_iB)\\&={\rm rank}z_1B+q-1,
		  	\end{align*}where comes from the fact that eigenvalues $\lambda_2,...,\lambda_q$ are controllable, leading to $z_iB\ne 0$ from the PBH test for $i=2,...,q$.
		  	Observe that %${\rm rank} [A,B]={\rm rank} [ZAZ^{-1},ZB]={\rm rank}[{\bf diag}\{\lambda_iI_{n_i}|_{i=1}^q\},{\bf col}\{z_i|_{i=1}^q\}B]= q\!-\!1+{\rm rank} z_1B$		  	
		  	{\begin{align*}&{\rm rank} [A,B]={\rm rank} [ZAZ^{-1},ZB]\\&\!=\!{\rm rank}[{\bf diag}\{\lambda_iI_{n_i}|_{i=1}^q\},{\bf col}\{z_i|_{i=1}^q\}B]\!=\! q\!-\!1+{\rm rank} z_1B.
\end{align*}}This leads to ${\rm rank} Q(A,B)={\rm rank} [A,B]$. %${\bf col}\{z_i|_{i=1}^q\}$ %${\rm rank} [A,B]={\rm rank} [ZAZ^{-1},ZB]={\rm rank}[{\bf diag}\{\lambda_iI_{n_i}|_{i=1}^q\},[z_1;\cdots;z_q]B]= q-1+{\rm rank} z_1B$.
		  	%	\begin{equation}\label{equal-indx} {\rm rank} Q(A,B)={\rm rank} [A,B].\end{equation}
		  	Since this relation holds for almost all realizations $(A,B)$ of $(\bar A, \bar B)$, we reach ${\rm grank}\, [\bar A,\bar B]= {\rm grank}\, Q(\bar A, \bar B)$.
		  	
		  	If not every state vertex is input-reachable in ${\mathcal G}(\bar A, \bar B)$, without losing generality, assume that $[\bar A, \bar B]$ has the form as (\ref{reducible-form}). From Proposition \ref{SCC-relation}, since $\bar A$ is generically diagonalizable, so is $\bar A_{11}$. {Using the above result on $(\bar A_{11}, \bar B_1)$, we have ${\rm grank}\, [\bar A_r, \bar B]= {\rm grank}\,[\bar A_{11},\bar B_1]={\rm grank}\,Q(\bar A_{11},\bar B_1)={\rm grank}\, Q(\bar A, \bar B)$, where the last equality follows the same line as (\ref{partition-equality}).} The remaining statement is immediate from Theorem~\ref{solvable-index}.
		  \end{proof}
		
		  \begin{remark}By inspection, there is a linking of size $p$ from $X^2\cup U^2$ to $Y^0$ in ${\mathcal D}(\bar A_r, \bar B, \bar C)$, if and only if there is a subset $X_S\subseteq X$ with $|X_S|=p$ such that ${\rm grank}\,[\bar A_r, \bar B](X_S,:)=p$ and ${\rm grank}\,\bar C(:,X_S)=p$. While a similar condition is given in \citet[Theo 2]{li2020structural} for the SOC of undirected networks with the symmetric weight constraint $A=A^\intercal$, our criteria are applied to different system classes. Notably, based on Corollary \ref{corollay-symmetry}, by comparing \citet[Theo 2]{li2020structural} and Corollary \ref{corollay-diagonalizable-solvable}, {we conclude that an undirected network with the symmetric weight constraint is SOC, if and only if it is SOC without such a constraint.}  {The dilation-based structural target controllability condition presented in \citet[Theo 2]{montanari2023target} is equivalent to Corollary \ref{corollay-diagonalizable-solvable} when each row of $\bar C$ has at most one nonzero entry}. However, they have not defined generic diagonalizability, and their utilization of the diagonalizability condition is implicit.  Besides,  they require that each column of $\bar B$ contains at most one nonzero entry, which is unnecessary here. %(thus not making clear the class of systems this condition applies to) %In addition, the linking condition in Corollary 4 is more general than the dilation-based one when some rows of $\bar C$ have more than one nonzero entries.

%we conclude that the SOC conditions for undirected networks are identical, regardless of whether the symmetric weight constraint is imposed.
		  	
		  	%Compared to \cite{montanari2023target}, we have defined SFO differently. The classes of systems enabling efficient SOC verification are also quite different. Moreover, they consider structural target controllability, which is a special case of SOC by restricting that every row of the output matrix has only one nonzero entry.
		  	
		  	%given in \cite{li2020structural}.
		  	%				The condition in Corollary \ref{corollay-diagonalizable-solvable} is similar to the SOC condition of undirected networks with the symmetric weight constraint $A=A^\intercal$ in \cite{li2020structural}. However, they are applied to different classes of systems. In particular, from Corollary \ref{corollay-symmetry}, we know for undirected networks without the symmetric weight constraint, the SOC condition coincides with the one given in \cite{li2020structural}.   %Corollary \ref{corollay-diagonalizable-solvable} extends the SOC criterion of undirected networks with symmetric weight constraint \cite{li2020structural} to generically diagonalizable systems (without the symmetry constraint).
		  \end{remark}

	{ {\begin{remark}[Duality between SFO and SOC]
Although there is no direct duality between SFO and SOC \citep{iudice2019node} (i.e., the SFO of $(\bar A,\bar C,\bar F)$ is generally not equivalent to the SOC of $(\bar A^{\intercal}, \bar C^{\intercal}, \bar F)$), \cite{montanari2023target} introduced some weak and strong dualities between them. Here, ``weak" duality means that one property indicates the other, but the reverse may not hold, while ``strong" duality means that one property implies the other and vice versa under certain conditions. In this context, conditions for one property may lead to necessary or sufficient conditions for the other. For example, the SFO of $(\bar A,\bar C,\bar F)$ implies the SOC of $(\bar A^{\intercal}, \bar C^{\intercal}, \bar F)$ \citep[Theo 4]{montanari2023target}, although the inverse may not hold. For more details, see \cite{montanari2023target,montanari2024duality,montanari2024popov}.
	  	\end{remark}}}
%,montanari2024duality
%	  	 Although there is no direct duality between SFO and SOC \cite{iudice2019node} (that is, the SFO of $(\bar A,\bar C,\bar F)$ is generically not equivalent to the SOC of $(\bar A^{\intercal}, \bar C^{\intercal}, \bar F)$), \cite{montanari2023target} introduced some weak and strong duality between them (here, ``weak'' means one property indicates the other but the inverse may not hold, and ``strong'' means one property implies the other and vice versa under certain conditions).  In this vein, conditions for one property may lead to some necessary/sufficient conditions for the other one. For example, the SFO of $(\bar A,\bar C,\bar F)$ implies the SOC of $(\bar A^{\intercal}, \bar C^{\intercal}, \bar F)$ (while the inverse may not hold). For more details, see \cite{montanari2023target,montanari2024popov}.
		
      	\section{The minimal sensor and actuator placement problems} \label{sec-sensor}		

      %\subsection{The minimal sensor placement problem}
      \subsection{Minimal sensor placement for SFO}
      The minimal sensor placement problem for SFO can be formulated as
      \begin{equation}\begin{array}{l}
      		\min \limits_{\bar C\in {\{0,*\}}^{p\times n}} p \tag{${\mathcal P}_1$}\label{prob1} \\
      		{\rm s.t.} \ (\bar A,\bar C,\bar F) \ {\rm SFO.}
      \end{array}\end{equation}
      Problem ${\mathcal P}_1$ (${\mathcal P}_1$ in short) asks to minimize the number of linear functions of states needed to be measured to estimate $z(t)$ ($z(t)$ is a linear combination of state variables with generic coefficients). Notice that there is no constraint on the structure of the available $\bar C$. {It has been shown in \citet[Prop 5]{zhang2023functional}} that if prior structure constraint is imposed on $\bar C$, such as the dedicated output constraint (i.e., each sensor measures only one state variable), ${\mathcal P}_1$ is NP-hard even for systems with self-loops in all states. This means even within the class of generically diagonalizable systems, imposing prior structure constraint to $\bar C$ will lead to the NP-hardness of ${\mathcal P}_1$. Nevertheless, the unconstrained case remains open.

{\begin{remark}[Non-monotonicity of SFO]
      It is noteworthy that the challenge of ${\mathcal P}_1$ lies in the `non-monotonicity' of SFO w.r.t. edge additions, that is, adding additional edges from state vertices to the existing sensors may destroy SFO. To illustrate this, consider two triples $(\bar A, \bar C_i, \bar F)$, $i\in \{1,2\}$, in which $\bar A= 0_{2\times 2}$, $\bar C_1=[*,0]$, $\bar C_2=[*,*]$, and $\bar F=[*,0]$. ${\mathcal G}(\bar A, \bar C_2)$ is obtained from ${\mathcal G}(\bar A, \bar C_1)$ by adding an edge from the second state vertex to the unique sensor. It can be verified that $(\bar A, \bar C_1, \bar F)$ is SFO while $(\bar A, \bar C_2, \bar F)$ not.
\end{remark}}
       %Such non-monotonicity makes ${\mathcal P}_1$ more complicated than the minimal sensor placement problem for structural observability \cite{Y.Y.2011Controllability}.\footnote{This is because SFO is defined from a generic perspective; see Remark \ref{example-counter}.}

      Theorem \ref{main-theorem} implies that the SFO of $(\bar A, \bar C, \bar F)$ for a generically diagonalizable system depends only on $X_F$, i.e., the nonzero columns of $\bar F$. To derive solutions for the minimal sensor problem, we first generalize this result to general systems, shown as follows.%\footnote{A similar result has also been observed in \cite{zhang2023functional}. Corollary \ref{single-pro} is presented here for self-consistency.} %. Corollary \ref{single-pro} is presented here for further derivations.

      \begin{corollary} \label{single-pro}
      	%	The triple $(\bar A, \bar C, \bar F)$ is SFO, if and only if $(\bar A, \bar C, \bar I_{X_F})$ is SFO.
      	Given a triple $(\bar A, \bar C, \bar F)$,	the following statements are equivalent:
      	\begin{itemize}
      		\item[(a)] $(\bar A, \bar C, \bar F)$ is SFO.
      		\item[(b)] $(\bar A, \bar C, \bar I_{X_F})$ is SFO.
      		\item[(c)] $(\bar A, \bar C, \bar e_i)$ is SFO for each $x_i\in X_F$.
      		\item [(d)] $(\bar A, \bar C, \bar I_{X_s})$ is SFO for every non-empty $X_s\subseteq X_F$.
      	\end{itemize}
      \end{corollary}

      \begin{proof} By regarding $O(\bar A, \bar C)$ as $[\bar A;\bar C]$ in the proof of `(b)$\Leftrightarrow(c)$' of Theorem \ref{main-theorem}, we can establish that ${\rm grank}\,[O(\bar A, \bar C);\bar F]={\rm grank}\,[O(\bar A, \bar C)]$, if and only if ${\rm grank}\,[O(\bar A,\bar C)]=$ ${\rm grank}\,[O(\bar A, \bar C);\bar e_i]$ for each $x_i\in X_F$.
      	The result then follows from (ii) of Lemma \ref{characterization-pro}.
      \end{proof}

      %This is due to the fact that SFO is defined from a generic perspective but cannot be defined as the property of allowing a functionally observable realization.
      %(and most minimal sensor placement problems for other properties of structured systems)

      %In this subsection, we give a closed-form solution to ${\mathcal P}_1$ within the class of generically diagonalizable systems. For general systems, we identify a non-decreasing property and based on which, propose two algorithms to obtain an upper bound for ${\mathcal P}_1$ in polynomial time, proven optimal under certain conditions.  %in the generically diagonalizable case or the case $\bar F=\bar I_n$. %, as well as graph-theoretic characterizations on the pertinent sensor positions.

      \subsubsection{Generically diagonalizable system case}
      We give a closed-form solution as well as a weighted maximum matching based algorithm for ${\mathcal P}_1$ in the class of generically diagonalizable systems.
      \begin{theorem}\label{optimal_sensor_theorem}
      	Given a pair $(\bar A,\bar F)$, suppose $\bar A$ is generically diagonalizable. Then, the optimal value $p^{*}$ to ${\mathcal P}_1$ is
      	\begin{equation} \label{optimal_value}
      		p^*={\max }\left\{{\rm grank}\,[\bar A; \bar I_{X_F}]-{\rm grank}\,\bar A, 1\right\}.
      	\end{equation}Moreover, let $X_{F}^u$ be a subset of $X_F$ with the maximum cardinality that is right-unmatched in a maximum matching of ${\mathcal B}(\bar A)$. Then, $p^*=\max\{|X_{F}^u|,1\}$, and Algorithm \ref{alg1} can find an optimal solution to~${\mathcal P}_1$.
      \end{theorem}

      \begin{proof} According to {Corollary \ref{single-pro}}, $p^*$ is the optimal value to ${\mathcal P}_1$ on $(\bar A, \bar F)$, if and only if  $p^*$ is the optimal value of ${\mathcal P}_1$ on $(\bar A, \bar I_{X_F})$. Suppose $\bar C_f$ is a feasible solution to ${\mathcal P}_1$. Theorem \ref{main-theorem} yields
      	$ {\rm grank}\,[\bar A;\bar C_f; \bar I_{X_F}]={\rm grank}\,[\bar A;\bar C_f]. $
      	This implies
      	${\rm grank}\,[\bar A;\bar C_f]\ge {\rm grank}\,[\bar A;\bar I_{X_F}].$
      	Since ${\rm grank}\,[\bar A;\bar C_f]\le {\rm row}(\bar C_f)+{\rm grank}\,\bar A$, we have
      	${\rm row}(\bar C_f)\ge {\rm grank}\,[\bar A;\bar I_{X_F}]- {\rm grank}\,\bar A,$
      	which indicates (\ref{optimal_value}) is a lower bound of ${\mathcal P}_1$.
      	
      	We are to show that the lower bound (\ref{optimal_value}) can be achieved via Algorithm \ref{alg1}.	By the construction of the weighted bipartite graph ${\mathcal B}(\bar A)$, it is obvious that $X_F^u$ is a subset of $X_F$ with the maximum cardinality that is right-unmatched in a maximum matching of ${\mathcal B}(\bar A)$. We now show that $|X_F^u|={\rm grank}\,[\bar A; \bar I_{X_F}]-{\rm grank}\,\bar A$. To show this,  from the property of rank function (cf. \citet[Prop. 2.1.3]{Murota_Book}), we know that there exist $S_1\subseteq [n]$ and $X_2\subseteq X_F$ satisfying ${\rm grank}\,\bar A(S_1,X\backslash X_2)=|S_1|={\rm grank}\,\bar A$, ${\rm grank}\, \bar I_{X_F}(:,X_2)=|X_2|$ such that ${\rm grank}\,[\bar A;\bar I_{X_F}]\!=\!{\rm grank}\,\bar A(S_1,:)\!+\!{\rm grank}\, \bar I_{X_F}(:,X_2)\!=\!{\rm grank}\,\bar A(S_1,X\backslash X_2)\!+\!{\rm grank}\, \bar I_{X_F}(:,X_2)\!=\!|S_1|+|X_2|$. As a result, ${\rm grank}\,[\bar A; \bar I_{X_F}]-{\rm grank}\,\bar A=|X_2|$ and $X_2\subseteq X_F$ is right-unmatched in some maximum matching of ${\mathcal B}(\bar A)$ (noting that $\bar A(S_1,X\backslash X_2)$ corresponds to a maximum matching of ${\mathcal B}(\bar A)$). Suppose $X_2$ is not of the maximum cardinality over all such possible sets, that is, there is a maximum matching ${\mathcal M}'$ of ${\mathcal B}(\bar A)$ associated with which the set of right-unmatched vertices belonging to $X_F$, denoted by $X_2'$, has a larger size than $|X_2|$. Then, ${\rm grank}\,[\bar A;\bar I_{X_F}]\ge |S_1|+|X_2'|>|S_1|+|X_2|$, since adding $|X_2'|$ edges to ${\mathcal M}'$ will form a new matching in ${\mathcal B}([\bar A;\bar I_{X_F}])$ with size $|S_1|+|X_2'|$, contradicting the fact that ${\rm grank}\,[\bar A;\bar I_{X_F}]=|S_1|+|X_2|$. %By regarding $X_2$ as $X_F^u$, the claim follows from Theorem \ref{optimal_sensor_theorem}.

       Let $\bar C'$ denote the matrix obtained after Step 3, and $\bar C$ the matrix obtained via Algorithm \ref{alg1}. Consider the case ${\rm grank}\,[\bar A; \bar I_{X_F}]-{\rm grank}\,\bar A\ge 1$. From the above analysis, we have ${\rm grank}\,[\bar A; \bar C']={\rm grank}\,[\bar A;\bar I_{X_F}]={\rm grank}\,\bar A+ p^*$. Notice that each row of $\bar C'$ is also a row of $\bar I_{X_F}$ with only one nonzero entry, leading to  ${\rm grank}\,[\bar A; \bar C';\bar I_{X_F}]={\rm grank}\,[\bar A; \bar I_{X_F}]$. Hence, ${\rm grank}\,[\bar A; \bar C']={\rm grank}\,[\bar A; \bar C';\bar I_{X_F}]$.
      	Let $X_{F}^c=X\backslash X_F$. Observe that all nonzero columns of $\bar C$ correspond to state vertices contained in $X_F$, yielding
      	${\rm grank}\,[\bar A; \bar C; \bar I_{X_F}]=|X_F|+ {\rm grank}\, \bar A_{X_F^c},$
      	where $\bar A_{X_F^c}\doteq \bar A(:,X_F^c)$.
      	On the other hand, by construction, it holds that
      	%	$${\rm grank}\,\left[\begin{array}{c}
      		%		\bar A\\
      		%		\bar C'
      		%	\end{array}\right]\le {\rm grank}\,\left[\begin{array}{c}
      		%		\bar A\\
      		%		\bar C
      		%	\end{array}\right]\le |X_F|+ {\rm grank}\, \bar A_{X_F^c}.$$
      	${\rm grank}\,[\bar A;\bar C']\le {\rm grank}\,[\bar A;\bar C]\le |X_F|+ {\rm grank}\, \bar A_{X_F^c}.$
      	Since
      	${\rm grank}\,[\bar A; \bar C']={\rm grank}\,[\bar A; \bar I_{X_F}]=|X_F|+ {\rm grank}\, \bar A_{X_F^c}$, it follows that ${\rm grank}\,[\bar A;\bar C]=|X_F|+ {\rm grank}\, \bar A_{X_F^c}$. We therefore reach
      	% $${\rm grank}\,\left[\begin{array}{c}
      		% 	\bar A\\
      		% 	\bar C\\
      		% 	\bar F
      		% \end{array}\right]=|X_F|+ {\rm grank}\, \bar A_{X_F^c}= {\rm grank}\,\left[\begin{array}{c}
      		% 	\bar A\\
      		% 	\bar C
      		% \end{array}\right].$$
      	${\rm grank}\,[\bar A; \bar C; \bar I_{X_F}]=|X_F|+ {\rm grank}\, \bar A_{X_F^c}= {\rm grank}\,[\bar A;\bar C].$
      	As all $x_i\in X_F$ are output-reachable in ${\mathcal G}(\bar A, \bar C)$, Theorem \ref{main-theorem}
      	indicates that $(\bar A, \bar C, \bar F)$ is SFO.
      	If ${\rm grank}\,[\bar A; \bar I_{X_F}]-{\rm grank}\,\bar A=0$, since all nonzero columns of $\bar C$ correspond to exactly $X_F$, we get ${\rm grank}\,[\bar A;\bar C; \bar I_{X_F}]=|X_F|+ {\rm grank}\, \bar A_{X_F^c}={\rm grank}\,[\bar A; \bar I_{X_F}]$. As ${\rm grank}\,[\bar A; \bar I_{X_F}]={\rm grank}\,\bar A$, it holds ${\rm grank}\,[\bar A;\bar C]={\rm grank}\,\bar A={\rm grank}\,[\bar A;\bar C; \bar I_{X_F}]$. As all $x_i\in X_F$ are output-reachable, $(\bar A, \bar C, \bar F)$ is SFO.
      \end{proof}

      \begin{algorithm}[H]
      	{{{
      				\caption{: A weighted maximum matching based algorithm for ${\mathcal P}_1$ in generically diagonalizable systems}  %to find an optimal solution to
      				\label{alg1}
      				\begin{algorithmic}[1]
      					\Require $\bar A$ and $\bar F$ with $\bar A$ generically diagonalizable.
      					\Ensure  An optimal solution to ${\mathcal P}_1$.
      					\State Construct a weighted bipartite graph ${\mathcal B}(\bar A)$ by assigning edge cost $w(e):E_{XX}\to \{0,1\}$ as
      					\begin{align}\label{cost}
      						\begin{split}	
      							w(e)= \left\{			
      							\begin{array}{cc}
      								1,                    & {\rm if} \ e=(x_i,x_j), x_i\in X_F\\
      								0,                     & {\rm otherwise.}
      							\end{array}
      							\right.
      						\end{split}
      					\end{align}
      					\State Find a minimum weight maximum matching ${\mathcal M}$ of ${\mathcal B}(\bar A)$. Let $X_{\mathcal M}=\{x_i: (x_i,x_j)\in {\mathcal M}\}$ be the set of right-matched vertices in ${\mathcal M}$, $X_S=X_F\cap  X_{\mathcal M}$, and $X_F^u=X_F\backslash X_{S}$. %Suppose that $X_S$ are located at $q$ different SCCs of ${\mathcal G}(\bar A)$.
      					\State Construct a $\max\{1,|X_F^u|\}\times n$ matrix $\bar C$ such that $\bar C_{ki}=*$ if $x_i$ is the $k$th element in $X_F^u$, $k=1,...,|X_F^u|$.
      					\State Per $x_i\in X_S$, choose one $j\in \{1,...,\max\{1,|X_F^u|\}\}$ and let $\bar C_{ji}=*$.
      					\State Return $\bar C$.% with ${\rm row}(\bar C)=p^*$.								
      					%								\If{${\rm grank}\,[\bar A; \bar I_{X_F}]-{\rm grank}\,\bar A\ge 1$}
      					%								\While{${\rm grank}\,[\bar A;\bar I_{X_F}]-{\rm grank}\,[\bar A;\bar C]>0$}
      					%								\State Find a $\bar e_i$ with $x_i\in X_F$ such that ${\rm grank}\,[\bar A;\bar C ;\bar e_i]-{\rm grank}\,[\bar A;\bar C]=1$.
      					%								\State Update $\bar C=[\bar C;\bar e_i]$.
      					%								\EndWhile
      					%								\EndIf						
      					%								\State Determine the output-unreachable state subset $X_S\subseteq X_F$ in ${\mathcal G}(\bar A, \bar C)$.
      					%								\State Let $\bar C_{1i}=*$, $\forall x_i\in X_S$.
      					%								\State Return $\bar C$ with ${\rm row}(\bar C)=p^*$.
      		\end{algorithmic}}}
      	}
      \end{algorithm}

            \begin{remark} From the proof of Theorem \ref{optimal_sensor_theorem}, it is easy to see that in Step 4 of Algorithm \ref{alg1}, instead of making $\bar C_{ji}=*$ for each $x_i\in X_S$, it suffices to choose those $x_i\in X_S$,  making $\bar C_{ji}=*$, such that every $x_i\in X_S$ is output-reachable in the resulting ${\mathcal G}(\bar A, \bar C)$.
      \end{remark} %only state vertices in $X_S$ that belong to a source-SCC of ${\mathcal G}(\bar A)$ need to be considered (i.e., making $\bar C_{1i}=*$ for the corresponding state $x_i$). Moreover, %	if a set of state vertices $X_S'\subseteq X_S$ belong to the same source-SCC, then only (and arbitrarily) one of them needs to be directly linked to a sensor.
      %   provided that the vertices of $X_S$ belong to $r$ different SCCs of ${\mathcal G}(\bar A)$, we can randomly pick one $x_i\in X_S$ per SCC and make $\bar C_{ji}=*$. %Furthermore, we can alternatively pick one $x_i\in X_S$ from each possible {\emph{sink-SCC}} of ${\mathcal G}(\bar A)$ to make $\bar C_{1i}=*$ (a sink-SCC is an SCC without incoming edges from other SCCs), and after that, randomly pick one $x_i\in X_S$ from the remaining SCCs that are not output-reachable in the resulting ${\mathcal G}(\bar A, \bar C)$. This may reduce the sparsity of $\bar C$ (i.e., number of nonzero entries).
      \begin{remark}
      	When $\bar F=\bar I_n$, Theorem \ref{optimal_sensor_theorem} collapses to the maximum matching based solution to the minimal sensor placement problem for structural observability (cf. \cite{Y.Y.2011Controllability}).
      \end{remark}

     { An example of the procedure of Algorithm \ref{alg1} is given in Fig. \ref{alg1-example}.} Notably, if $\bar A$ is generically non-diagonalizable, (\ref{optimal_value}) only gives a lower bound for ${\mathcal P}_1$ (see Example \ref{sensor-placement-example}). %This can be seen from the following example.

%      \begin{example} \label{exampl1}
%      	Consider a system whose associated ${\mathcal G}(\bar A)$ is given in Fig. \ref{sensor-placement-example}(a).
%      	%	Consider the system $(\bar A, \bar F)$ given by
%      	%				{\small{$$\bar A=\left[\begin{array}{cccccc}
%      				%							0 & 0 & 0 & 0 & 0 & 0\\
%      				%							* & 0 & 0 & 0 & 0 & 0\\
%      				%							0 & * & 0 & 0 & 0 & 0\\
%      				%							0 & 0 & 0 & 0 & 0 & 0\\
%      				%							0 & * & 0 & * & 0 & 0 \\
%      				%							0 & * & 0 & * & 0 & 0\\
%      				%						\end{array}
%      			%						\right],\bar F=[0,*,*,*,0,0].$$}}
%      	%	The digraph ${\mathcal G}(\bar A)$ is given in Fig. \ref{sensor-placement-example}(a).
%      	It turns out that $\bar A$ is generically non-diagonalizable, and
%      	${\rm grank}\,[\bar A; \bar I_{X_F}]-{\rm grank}\,\bar A=1$, resulting in $p^*=1$ (see Fig. \ref{sensor-placement-example}(a) for the maximum matching based interpretation). However, it can be verified via Lemma \ref{characterization-pro} that,
%      	for any $\bar C\in \{0,*\}^{1\times 6}$, $(\bar A, \bar C, \bar F)$ is not SFO. %This illustrates that for generically non-diagonalizable systems, $p^*$ given in Theorem \ref{optimal_value} may only be a lower bound.
%      \end{example}

   \begin{figure}[H]
      	\centering
      	% Requires \usepackage{graphicx}
      	\includegraphics[width=2.9in]{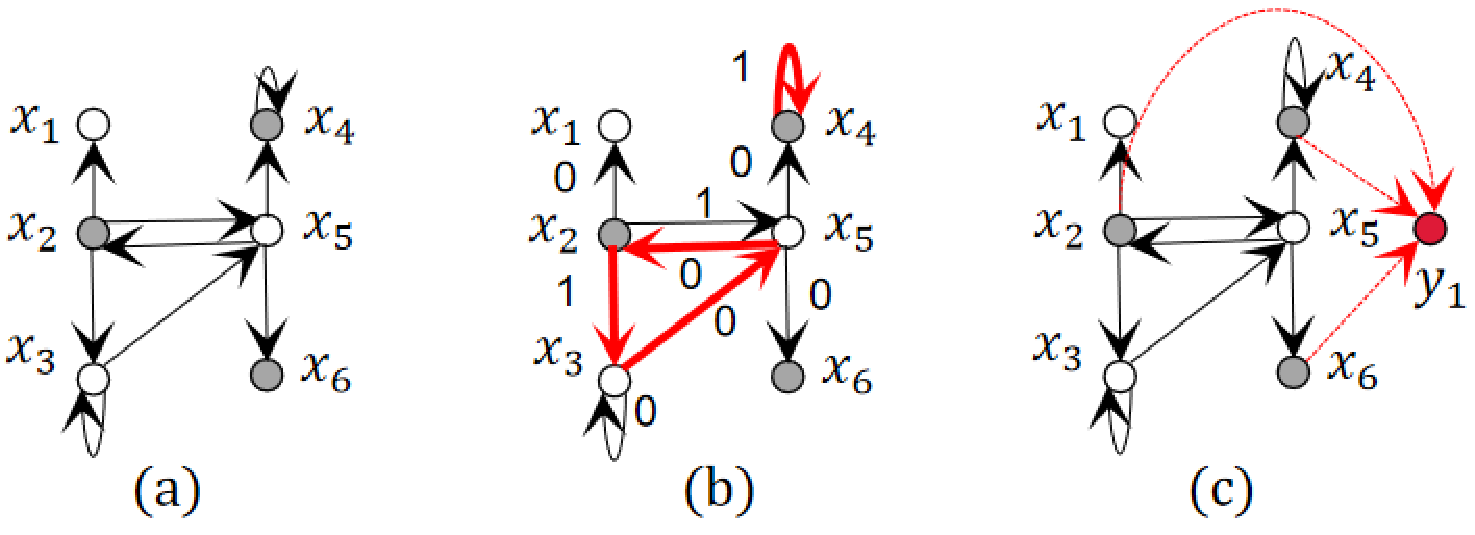}\\
      	\caption{  { Example of the procedure of Algorithm \ref{alg1} applied to the system digraph ${\mathcal G}(\bar A)$ in (a), where $\bar A$ is generically diagonalizable. Nodes in gray represent functional states. In (b), the number on each edge is the cost assigned to it (see (\ref{cost})), and bold red edges form a minimum weight maximum matching ${\mathcal M}$ of ${\mathcal B}(\bar A)$. Accordingly, $X_S=\{x_2,x_4\}$ and $X_F^u=\{x_6\}$, resulting in $p^*=1$.  In (c), the dotted red edges represent the obtained solution.} }\label{alg1-example}
      \end{figure}

      \subsubsection{General case}
      %			In this subsection, we propose two algorithms to obtain an upper bound for ${\mathcal P}_1$ in the general case, proven optimal under certain conditions.  %give an upper bound for ${\mathcal P}_1$ in the general case (i.e., without the generic diagonalizability condition).
      %			Our bound is based on a non-decreasing property of SFO w.r.t. a class of edge additions.

      In this subsection, we propose two algorithms to obtain an upper bound for ${\mathcal P}_1$ in the general case, based on a non-decreasing property of SFO w.r.t. a class of edge additions. {This bound is proven optimal under certain circumstances.}

      %To ease the descriptions, we first introduce some notations.d

    Given an output matrix $\bar C\in \{0,*\}^{p\times n}$, denote $X_C$ as the set of state vertices corresponding to the nonzero columns of $\bar C$ (that is, $X_C$ is defined similarly to $X_F$), and denote $E_{C}=\{(x_i,y_j): \bar C_{ji}\ne 0\}$ as the set of edges corresponding to nonzero entries of $\bar C$.     {Recall that ${\mathcal G}(\bar A, \bar C)=(X\cup Y,E_{XX}\cup E_{XY})$. Let $d({\mathcal G}(\bar A, \bar C))$ be the maximum size of an output cactus configuration in $d({\mathcal G}(\bar A, \bar C))$, i.e., the maximum number of output-reachable state vertices in a collection of disjoint output stems and cycles of ${\mathcal G}(\bar A, \bar C)$. Hereafter, ``output'' will be dropped from ``output cactus configuration'' for simplicity if no confusion is made.  A {\emph{maximum cactus configuration}} refers to a cactus configuration of ${\mathcal G}(\bar A, \bar C)$ with the largest size.} {For a state vertex $x_i\in X$ and output vertex $y_j$, let ${\mathcal G}(x_i,y_j)$ be the graph consisting of an edge $(x_i,y_j)$ from $x_i$ to $y_j$ and the corresponding end vertices $\{x_i,y_j\}$, i.e., ${\mathcal G}(x_i,y_j)=(\{x_i,y_j\},(x_i,y_j))$. Given two digraphs ${\mathcal G}_1=(V_1,E_1)$ and ${\mathcal G}_2=(V_2,E_2)$, ${\mathcal G}_1\cup {\mathcal G}_2$ denotes the digraph $(V_1\cup V_2, E_1\cup E_2)$.} {The following lemma follows directly from \citet[Theo 1]{hosoe1980determination} and the duality between controllability and observability.}

%
% Denote by $d({\mathcal G}(\bar A, \bar C))$ the maximum size of a
%cactus configuration of ${\mathcal G}(\bar A, \bar C)$.
%Let $Y_C=\{y_1,...,y_{{\rm row}(\bar C)}\}$ be the set of sensors corresponding to $\bar C$.

    %  When ${\mathcal G}$ is a system digraph of a pair $(\bar A, \bar C)$,  let $d({\mathcal G})$ be the maximum number of state vertices that can be covered by a cactus configuration of ${\mathcal G}$. For two digraphs ${\mathcal G}_1=(V_1,E_1)$ and ${\mathcal G}_2=(V_2,E_2)$, ${\mathcal G}_1\cup {\mathcal G}_2$ denotes the digraph $(V_1\cup V_2, E_1\cup E_2)$.

      \begin{lemma}[Theo 1, \citeauthor{hosoe1980determination},\citeyear{hosoe1980determination}] \label{hosoe-subspace}
      	{Given a structured matrix pair $(\bar A, \bar C)$, it holds that ${\rm grank}\,O(\bar A, \bar C)=d({\mathcal G}(\bar A, \bar C))$.}
      \end{lemma}

      \begin{lemma} \label{cactus-adding-criterion}
      	The triple $(\bar A, \bar C, \bar F)$ is SFO, if and only if $d({\mathcal G}(\bar A, \bar C))=d({\mathcal G}(\bar A, \bar C)\cup {\mathcal G}(x_i,y_{p+1}))$ holds for every $x_i\in X_F$, {i.e., adding an additional dedicated sensor $y_{p+1}$ to measure state vertex $x_i$ will not change the generic dimension of unobservable subspaces}.
      \end{lemma}
      \begin{proof} The result follows directly from Lemma \ref{characterization-pro} (note ${\rm grank}\,[O(\bar A,\bar C);O(\bar A,\bar F)]={\rm grank}\,O(\bar A,[\bar C;\bar F])$), Corollary \ref{single-pro}, and Lemma \ref{hosoe-subspace}. %and the relation ${\rm grank}\,O(\bar A, \bar C)=d({\mathcal G}(\bar A, \bar C))$.
      \end{proof}

      \begin{proposition} \label{non-decreasing}
      	%	If $(\bar A, \bar C, \bar F)$ is SFO and $X_C\subseteq X_F$, then $(\bar A, \bar C', \bar F)$ is SFO for any $\bar C'$ such that $E_{C}\subseteq E_{C'}$ and $X_{C'}\subseteq X_F$.
      	If $(\bar A, \bar C, \bar F)$ is SFO, then $(\bar A, \bar C', \bar F)$ is SFO for any $\bar C'$ such that $E_{C}\subseteq E_{C'}$ and ${\rm ini}(\Delta E)\subseteq X_F$, where $\Delta E\doteq E_{C'}\backslash E_C$, and ${\rm ini}(\Delta E)$ is the set of initial vertices of edges in $\Delta E$.
      \end{proposition}

{\begin{proof}
	To make this section more centralized, the proof is given in the Appendix.
\end{proof}}

      %\end{remark}	

      Despite the non-monotonicity of SFO w.r.t. edge additions, Proposition \ref{non-decreasing} reveals that adding edges from functional states to existing sensors would not destroy SFO. We call such a property the {\emph{non-decreasing property of SFO w.r.t. functional state measurement addition}}. Based on this property, in what follows, we provide two methods to determine an upper bound for ${\mathcal P}_2$ for general systems, which turns out to be optimal for ${\mathcal P}_2$ subject to the constraint $X_C\subseteq X_F$. % As a result, this upper bound becomes optimal under certain circumstances.

      % Algorithm \ref{alg2} finds a feasible solution to ${\mathcal P}_1$, which gives an upper bound for ${\mathcal P}_2$ for general systems.

      \begin{algorithm}[H]
      	{{{
      				\caption{: A naive iterative algorithm to determine a feasible solution to ${\mathcal P}_1$ for general systems}
      				\label{alg2}
      				\begin{algorithmic}[1]
      					%	\Require $\bar A$ and $\bar F$.
      					%	\Ensure  A feasible solution to ${\mathcal P}_1$.
      					\State {Set $\bar C=\emptyset$}
      					\While{${\rm grank}\,O(\bar A,[\bar C;\bar F])-{\rm grank}\,O(\bar A, \bar C)\ge 1$}
      					\State Update $\bar C=[\bar C; \eta]$, with $\eta \in \{0,*\}^{1\times n}$ satisfying $\eta_i=*$ if and only if $x_i\in X_F$.
      					\EndWhile
      					\State Return $\bar C$.
      		\end{algorithmic}}}
      	}
      \end{algorithm}

      \begin{algorithm}[H]
      	{{{
      				\caption{: A weighted maximum matching based algorithm for a feasible solution to ${\mathcal P}_1$ for general systems}
      				\label{alg3}
      				\begin{algorithmic}[1]
      					%	\Require $\bar A$ and $\bar F$.
      					%	\Ensure  A feasible solution to ${\mathcal P}_1$.
      					\State {Construct the weighted bipartite graph $\tilde {\mathcal B}(\bar A, \bar I_{X_F})$}
      					
      					\State Find the maximum weighted maximum matching ${\mathcal H}^*$ of  $\tilde {\mathcal B}(\bar A, \bar I_{X_F})$. Let $X_{{\mathcal H}^*}=
      					\{x_i\in X: (x_i,y_j)\in {\mathcal H}^*,y_j\in Y\}$.
      					\State Construct a $\gamma\times n$ matrix $\bar C$ such that $\bar C_{ki}=*$ if $x_i$ is the $k$th element (in any order) of $X_{{\mathcal H}^*}$, $k=1,...,|X_{{\mathcal H}^*}|$, $\gamma\doteq {\max}\{1,|X_{{\mathcal H}^*}|\}$.
      					\For {$x_j\in X_F\backslash X_{{\mathcal H}^*}$}
      					\If{$x_j$ is not reachable to any $x_k\in X_{{\mathcal H}^*}$}
      					\State Choose one $i\in \{1,...,\gamma\}$ and let $\bar C_{ij}=*$;
      					\EndIf
      					\EndFor
      					\State Return $\bar C$.
      		\end{algorithmic}}}
      	}
      \end{algorithm}

      {\emph{Naive iterative algorithm:}} The first method is a naive iterative algorithm, shown in Algorithm \ref{alg2}. The basic idea is that, since $\bar I_{X_F}$ is a feasible solution to ${\mathcal P}_1$ (cf. Lemma \ref{characterization-pro}), there exists a $\bar C'$ satisfying ${\rm row}(\bar C')\le |X_F|$ and $X_{C'}\subseteq X_F$ such that $(\bar A, \bar C', \bar F)$ is SFO. By Proposition \ref{non-decreasing}, whenever there is a feasible solution ${\bar C'}$ to ${\mathcal P}_1$ satisfying
      $X_{C'}\subseteq X_F$, the following output matrix
      $$\bar C=\underbrace{[\eta;\eta;...;\eta]}_{{\rm row}(\bar C') \ {\rm times}}$$
      is also feasible for ${\mathcal P}_1$ ($\eta$ is defined in Algorithm \ref{alg2}). Therefore, Algorithm \ref{alg2} finds a feasible solution $\bar C$ to ${\mathcal P}_1$ with the minimum number of rows subject to the constraint that $X_{C}\subseteq X_F$.

      {\emph{Weighted maximum matching based algorithm:}}	To avoid the {\emph{iterative}} steps in Algorithm \ref{alg2}, in what follows, we provide a weighted maximum matching based algorithm to directly determine (i.e., without iterations) the upper bound obtained in Algorithm \ref{alg2}.  The key idea is to find the minimum number of output stems in a maximum cactus configuration of ${\mathcal G}(\bar A, \bar I_{X_F})=(X\cup Y, E_{XX}\cup E_{XY})$, with $Y=\{y_1,...,y_{q}\}$, $q\doteq |X_F|$, and $E_{XX}$, $E_{XY}$ defined in Section \ref{pre-sec}. To this end, we construct a weighted bipartite graph $\tilde {\mathcal B}(\bar A, \bar I_{X_F})=(X\cup Y, X\cup Y, \tilde E_{XY})$, where $\tilde E_{XY}=E_{XX}\cup E_{XY} \cup \{(v,v):v\in X\cup Y\}\cup \{(y,x): y\in Y, x\in X\}$. It turns out that $\tilde {\mathcal B}(\bar A, \bar I_{X_F})$ corresponds to a digraph that is obtained from ${\mathcal G}(\bar A, \bar I_{X_F})$ by adding a self-loop to each vertex of $X\cup Y$ and a return edge from each vertex of $Y$ to each one of $X$ (see Fig. \ref{sensor-placement-example}(c) for illustration). Let $W_F\subseteq X$ be the set of state vertices that are output-reachable in ${\mathcal G}(\bar A,\bar I_{X_F})$. The edge cost
      $w(e): \tilde E_{XY}\to {\mathbb N}$ is defined as follows:
      \begin{align} \label{edge_cost}
      	\begin{split}	
      		w(e)= \left\{			
      		\begin{array}{cc}
      			q+1,                    & {\rm if} \ e\in E_{XX}^F\\
      			q,                     & {\rm if} \ e\in E_{XY} \\
      			0,  & {\rm otherwise,}
      		\end{array}
      		\right.
      	\end{split}
      \end{align}where $E_{XX}^F\doteq \{(x_i,x_j)\in E_{XX}: x_j\in W_F\}$ is the set of state edges within $W_F$.
      Building on $\tilde {\mathcal B}(\bar A, \bar I_{X_F})$, Algorithm \ref{alg3} finds a feasible solution to ${\mathcal P}_1$ with the same number of sensors as Algorithm \ref{alg2} without relying on iterations; see the following theorem.

      \begin{theorem} \label{theorem-general-case}
      	Both Algorithms \ref{alg2} and \ref{alg3} can find a feasible solution to ${\mathcal P}_1$, which has the minimum number of rows over all feasible solutions to ${\mathcal P}_1$ subject to the constraint that $X_C\subseteq X_F$. Moreover, if $\bar A$ is generically diagonalizable or $\bar F= \bar I_n$, this solution is optimal for ${\mathcal P}_1$.
      	% an optimal solution to ${\mathcal P}_1$ subject to the constraint that $X_C\subseteq X_F$. Moreover, if $\bar A$ is generically diagonalizable or $\bar F= \bar I_n$, Algorithms \ref{alg2} and \ref{alg3} gives an optimal solution to ${\mathcal P}_1$.
      	
      	% Algorithm \ref{alg2} gives a feasible solution to ${\mathcal P}_1$. Moreover, if $\bar A$ is generically diagonalizable or $\bar F= \bar I_n$, Algorithm \ref{alg2} gives an optimal solution to ${\mathcal P}_1$.
      	
      	% Therefore, Algorithm \ref{alg2} finds a feasible solution $\bar C$ with the minimum number of rows subject to the constraint that $X_{C}\subseteq X_F$. Hence, the optimality of Algorithm \ref{alg2} when $\bar F=\bar I_n$ is obvious.
      	% 				Since the optimal solution $\bar C$ returned by Algorithm \ref{alg1} satisfies $X_{C}\subseteq X_F$, the solution given by Algorithm \ref{alg2} is also optimal for generically diagonalizable systems. %observe that the cactus configurations in ${\mathcal G}(\bar A, \bar I_{x_F})$ and maximum matchings of $\tilde {\mathcal B}(\bar A, \bar I_{X_F})$ are of one-one correspondence. To see this, $\tilde {\mathcal B}(\bar A, \bar I_{X_F})$,
      \end{theorem}

 { \begin{proof}
  	Due to its lengthiness, the proof is appended to the Appendix.
  \end{proof}}

      \begin{figure}
	\centering
	% Requires \usepackage{graphicx}
	\includegraphics[width=2.9in]{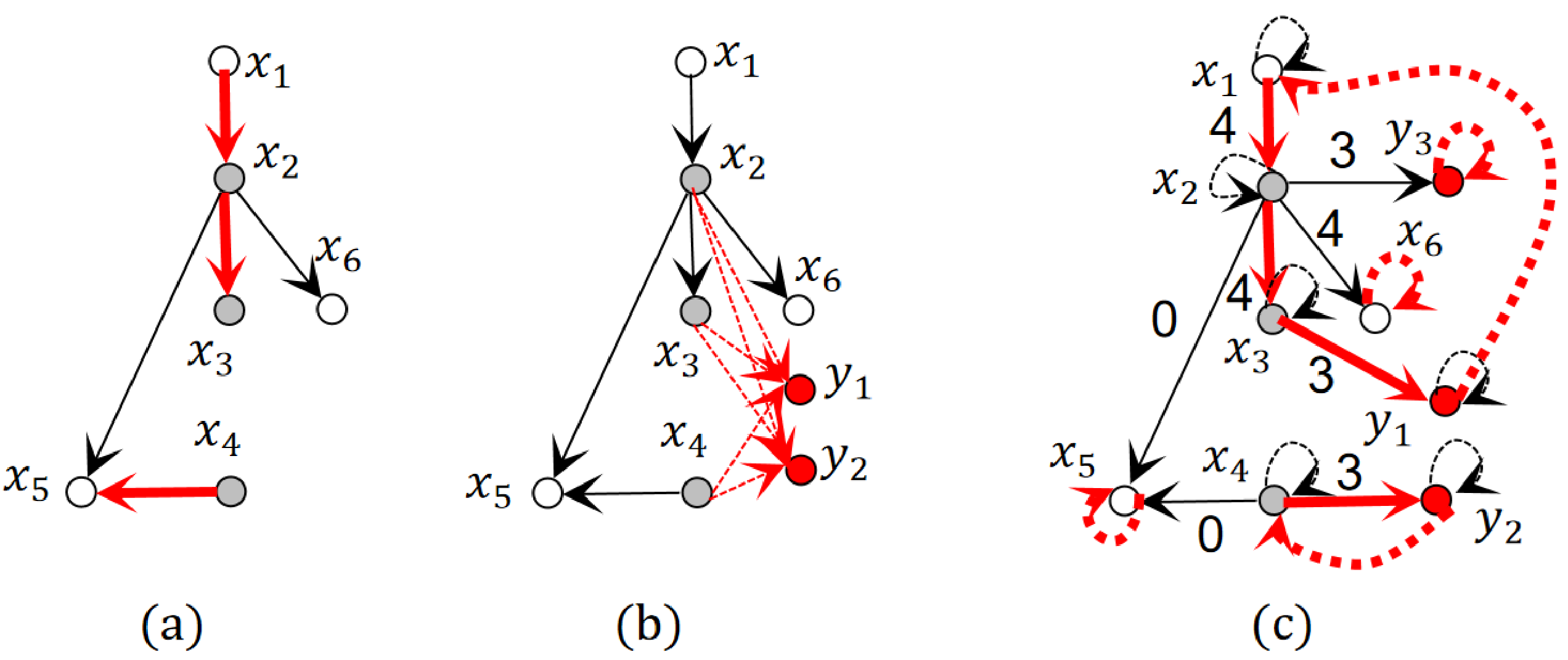}\\
	\caption{Illustration of Algorithms \ref{alg2} and \ref{alg3} applied to the system digraph ${\mathcal G}(\bar A)$ in (a). Nodes in gray represent functional states, and in red represent sensors. In (a),  bold red edges form a maximum matching ${\mathcal M}$ that does not right match $x_3\in X_F$ in ${\mathcal B}(\bar A)$, resulting in $p^*=1$ via Algorithm \ref{alg1}.  In (b), the dotted red edges represent the solution returned by Algorithm \ref{alg2}. { Subfigure (c) presents the digraph corresponding to the bipartite graph $\tilde {\mathcal B}(\bar A, \bar I_{X_F})$, where partial return edges from $Y$ to $X$ are omitted for simplicity. The number on each edge is the cost assigned to it via (\ref{edge_cost}), where zero costs are omitted for edges not belonging to $E_{XX}\cup E_{XY}$. Bold (solid and dotted) red edges correspond to a maximum weighted maximum matching ${\mathcal H}^*$ of $\tilde {\mathcal B}(\bar A, \bar I_{X_F})$ with weight $4+4+3+3=14$. Accordingly, $X_{{\mathcal H}^*}=\{x_3,x_4\}$, and there are two output stems $(x_1,x_2,x_3,y_1)$ and $(x_4,y_2)$ associated with it in ${\mathcal G}(\bar A, \bar I_{X_F})$. Algorithm \ref{alg3} then returns the solution of placing two dedicated sensors at $x_3,x_4$.} } \label{sensor-placement-example}
\end{figure}

%Notice that  In (d), the bold red edges form a maximum matching of ${\mathcal G}(\bar A)$. According to \cite{Y.Y.2011Controllability}, a minimal sensor set for structural observability is depicted by the dotted red edges.

%      	${\rm grank}\,[\bar A; \bar I_{X_F}]-{\rm grank}\,\bar A=1$, resulting in $p^*=1$ (see Fig. \ref{sensor-placement-example}(a) for the maximum matching based interpretation). However, it can be verified via Lemma \ref{characterization-pro} that,
%      	for any $\bar C\in \{0,*\}^{1\times 6}$, $(\bar A, \bar C, \bar F)$ is not SFO.

      \begin{example}\label{example-cont}
      	Consider the system given in Fig. \ref{sensor-placement-example} (a), where the corresponding $\bar A$ is generically non-diagonalizable. Algorithm \ref{alg1} applied to this system results in $p^*=1$ (see Fig. \ref{sensor-placement-example}(a) for the maximum matching based interpretation). However, it can be verified via Lemma \ref{characterization-pro} that,
      	for any $\bar C\in \{0,*\}^{1\times 6}$, $(\bar A, \bar C, \bar F)$ is not SFO. Now apply Algorithm \ref{alg2} to this system. It turns out that if $\bar C_1=\eta$ with $\eta=[0,*,*,*,0,0]$, then ${\rm grank}\,O(\bar A, [\bar C_1;\bar F])- {\rm grank}\,O(\bar A, \bar C_1)=1$. If $\bar C_2=[\eta;\eta]$, ${\rm grank}\,O(\bar A, [\bar C_2;\bar F])- {\rm grank}\,O(\bar A, \bar C_2)$ $=0$. Hence, Algorithm \ref{alg2} gives a solution $\bar C_2$ to ${\mathcal P}_1$ with two sensors (see Fig. \ref{sensor-placement-example}(b) for illustration). Notice that $(\bar A, \bar C_2)$ is not structurally observable. {Fig. \ref{sensor-placement-example}(c) presents a set of edges corresponding to a maximum weighted maximum matching in $\tilde {\mathcal B}(\bar A, \bar I_{X_F})$. Associated with it, there are two output stems in ${\mathcal G}(\bar A, \bar I_{X_F})$. Therefore, two dedicated sensors deployed on $x_3$ and $x_4$ can guarantee the SFO.} %, whose weight is $4+4+3+3=14$
      \end{example}
      %As shown in Example xx, this property does not hold when adding edges from the non-functional states to existing sensors.
      %The maximum matching based approach given in \cite{Y.Y.2011Controllability} yields that the minimal number of sensors to achieve structural observability is $6- {\rm grank}\,\bar A=3$ (see Fig. \ref{sensor-placement-example}(d))
      % \begin{proposition}
      	%  		If $(\bar A, \bar C, \bar F)$ is SFO, then there exists $\bar C'$ satisfying ${\rm row}(\bar C')\le {\rm row}(\bar C)$ and $X_{\bar C'}\subseteq X_F$, such that $(\bar A, \bar C', \bar F)$ is SFO.
      	%  \end{proposition}

      %	\begin{figure}
      	%	\centering
      	%	% Requires \usepackage{graphicx}
      	%	\includegraphics[width=3.0in]{SFO-optimal.eps}\\
      	%	\caption{${\mathcal G}(\bar A, \bar C)$ and its associated ${\mathcal D}(\bar A, \bar C)$ in Example \ref{counter-example-2}.} \label{SFO-example-digraph}
      	%\end{figure}
      	
      {	{\bf Conjecture:}}
      		We conjecture that the solution obtained by Algorithm \ref{alg3} is optimal for ${\mathcal P}_1$ even without the constraint $X_C\subseteq X_F$. {Our extensive examples consistently support this conjecture.  However, we are currently unable to provide a formal proof or find a counterexample. Addressing this conjecture remains an open problem for future research.}

    %  However,  we are currently unable to prove this conjecture (or find a counter example), which is left for future work. % We leave proving or disproving this conjecture for future work.
      %	\end{remark}
      	
      	\subsection{Minimal actuator placement for SOC} \label{soc-optimization}
      	The minimal actuator placement problem seeks to determine the minimum number of actuators (independent inputs), also known as driver nodes \citep{gao2014target}, to achieve SOC. Formally, given $(\bar A, \bar C)$, this problem is formulated as
      	\begin{equation}\begin{array}{l}
      			\min \limits_{\bar B\in {\{0,*\}}^{n\times m}} m \tag{${\mathcal P}_2$}\label{prob2} \\
      			{\rm s.t.} \ (\bar A,\bar B,\bar C) \ {\rm SOC.}
      	\end{array}\end{equation}The above problem is denoted by ${\mathcal P}_2$. In ${\mathcal P}_2$, we do not impose any structure constraint on $\bar B$. It has been shown in \citet[Theo 2]{czeizler2018structural} that determining the minimal actuators needed for SOC is NP-hard when the number of states that each actuator can simultaneously actuate is bounded by a constant $N$, for any $N\ge 1$. The unconstrained ${\mathcal P}_2$, however, remains open. In this subsection, we reveal that in the class of generically diagonalizable systems, ${\mathcal P}_2$ can be solved in polynomial time, in particular, via a weighted maximum flow algorithm.
      	
      	Given a directed graph (network) ${\mathcal G}=(V,E)$ with a source node $s$ and a sink node $t$, suppose each edge $e=(u,v)$ has a capacity $c(u,v)$ and a cost $w(u,v)$. A flow $f:E\to {\mathbb R}_{\ge 0}$ over ${\mathcal G}$ is a map that assigns a non-negative number $f(e)$ to each edge $e\in E$ subject to that the flow $f(e)$ on each edge $e$ cannot exceed the edge capacity $c(e)$, and the sum of flows into a node equals the sum of flows out of that node, unless the source $s$ and the sink $t$ \citep{Ahuja1993NetworkFT}. The value of a flow over ${\mathcal G}$ is the sum of flows passing from the source to the sink, and the cost of a flow $f$ is defined as $w(f)=\sum_{e\in E} f(e)w(e)$. An integral flow refers to a flow $f$ such that $f(e)$ is integral for every $e\in E$. The maximum flow of a network ${\mathcal G}$ is the largest possible flow over ${\mathcal G}$. A minimum cost maximum flow of ${\mathcal G}$ is a maximum flow with the smallest possible cost.
      	
      	\begin{theorem} \label{output-polynomial}
      		Given $(\bar A, \bar C)$ with $\bar A$ generically diagonalizable and ${\rm grank}\,\bar C=p$, the optimal value of ${\mathcal P}_2$ is
      		$$m^*=\max\{1,p-{\rm grank}\,\bar C\bar A\},$$
      		and Algorithm \ref{alg4} can determine the corresponding optimal solution $\bar B^*$ in $O(n^3)$ time.
      	\end{theorem}
      	\begin{proof}
      		Suppose $(\bar A, \bar B, \bar C)$ is SOC. From Corollary \ref{corollay-diagonalizable-solvable} and the proof of Theorem \ref{solvable-index}, it holds that ${\rm grank}\,\bar C[\bar A_r, \bar B]=p$. Noting ${\rm grank}\,\bar C[\bar A_r, \bar B]\le {\rm grank}\,\bar C[\bar A, \bar B]\le {\rm grank}\,\bar C\bar A+ {\rm grank}\,\bar C\bar B$, we have ${\rm row}(\bar B)\ge {\rm grank}\,\bar C\bar B\ge p-{\rm grank}\,\bar C\bar A$. This indicates $m^*$ is a lower bound for ${\mathcal P}_2$.
      		
      		By the construction in Algorithm \ref{alg4}, the maximum flow of the network ${\mathcal F}(\bar A, \bar C)$ is exactly $p$. From the integral flow theorem \citep{Ahuja1993NetworkFT}, the minimum cost integral maximum flow $f^*$ exists. Moreover, all integral maximum flows are of one-one correspondence to linkings with size $p$ from $U^2\cup X^2$ to $Y^0$ in ${\mathcal D}(\bar A, \bar I_n,\bar C)$. Since ${\rm grank}\,\bar C\bar A$ equals the maximum size of a linking from $X^2$ to $Y^0$ in ${\mathcal D}(\bar A, \bar I_n, \bar C)$ (\citet[Theo. 6]{papadimitriou1984simple}), we have $|X^{f_2^*}|={\rm grank}\,\bar C\bar A$ and $|X^{f_1^*}|=p-{\rm grank}\,\bar C\bar A$. In addition, Step 3 of Algorithm \ref{alg4} ensures that the initial vertex of each path in the linking of size $|X^{f_2^*}|$ from $X^2$ to $Y^0$ in ${\mathcal D}(\bar A, \bar B^*, \bar C)$ is input-reachable. Therefore, $(\bar A, \bar B^*, \bar C)$ is SOC, and $\bar B^*$ has the minimum number $m^*$ of columns. The $O(n^3)$ complexity comes from
      		the fact that a minimum cost maximum flow of a network ${\mathcal G}=(V,E)$ can be determined in $O(|V||E|)$ time \citep{Ahuja1993NetworkFT}.
      	\end{proof}

      	\begin{algorithm}[H]
      		{{{
      					\caption{: A weighted maximum flow based algorithm to find an optimal solution to ${\mathcal P}_2$ in generically diagonalizable systems}
      					\label{alg4}
      					\begin{algorithmic}[1]
      						\Require $\bar A$ and $\bar C$ with $\bar A$ generically diagonalizable and ${\rm grank}\,\bar C=p$.
      						\Ensure An optimal solution $\bar B^*$ to ${\mathcal P}_2$.
      						\State Construct a flow network ${\mathcal F}(\bar A, \bar C)$ from ${\mathcal D}(\bar A, \bar I_n, \bar C)$ as follows: duplicate each vertex $v$ of ${\mathcal D}(\bar A, \bar I_n, \bar C)$ with two vertices $v^i,v^o$ and an edge $(v^i,v^o)$, and for each edge $(v,w)$ of ${\mathcal D}(\bar A, \bar I_n, \bar C)$, replace it with an edge $(v^o,w^i)$; add to the resultant graph a source $s$, a sink $t$, and the incident edges $\{(s,v^i): v\in U^2\cup X^2\}\cup \{(w^o,t):w\in Y^0\}$. The edge capacity is set as $c(e)=1$ for each edge $e$ of ${\mathcal F}(\bar A, \bar C)$. The edge cost is set as $w(e)=1$ if $e\in \{(u_j^{o},x_j^{1i}):j=1,...,n\}$, and $w(e)=0$ otherwise.
      						\State Find a minimum cost integral maximum flow $f^*$ of ${\mathcal F}(\bar A, \bar C)$. Let $X^{f_1^*}=\{x_i\in X: f(u_{i}^{o},x_{i}^{1i})\ne 0\}$, $X^{f_2^*}=\{x_i\in X: f(x_{i}^{2i},x_{i}^{2o})\ne 0\}$, and $m^*=\max\{1, |X^{f_1^*}|\}$. Suppose that $X^{f_2^*}$ are located at $q$ different SCCs of ${\mathcal G}(\bar A)$, with the vertex set of the $i$th one being $X_i\subseteq X$.
      						\State Construct an $n\times m^*$ matrix $\bar B_*$ such that $\bar B^*_{jk}=*$ if $x_j$ is the $k$th element of $X^{f^*_1}$, $k=1,...,|X^{f_1^*}|$. In addition, per $i\in \{1,...,q\}$, choosing one $x_j\in X_i$ and one $k\in \{1,...,m^*\}$, let $\bar B^*_{jk}=*$.
      						\State Return $\bar B^*$.
      			\end{algorithmic}}}
      		}
      	\end{algorithm}
\begin{example}
	     {Consider a system $(\bar A, \bar C)$ whose ${\mathcal G}(\bar A, \bar C)$ is given in Fig. \ref{driver-placement-soc}(a).  It can be verified that $\bar A$ is generically diagonalizable. Fig. \ref{driver-placement-soc}(b) presents the corresponding flow network ${\mathcal F}(\bar A, \bar C)$ constructed in Algorithm \ref{alg4}. In the minimum cost maximum flow $f^*$ (edges with nonzero flows are highlighted in bold in Fig. \ref{driver-placement-soc}(b)), $X^{f_1^*}=\{x_2\}$ and $X^{f_2^*}=\{x_2,x_4\}$, yielding $m^*=1$. From step 3 of Algorithm \ref{alg4}, deploying a dedicated actuator at $x_2$, corresponding to $\bar B^*=[0,*,0,0,0]^{\intercal}$, suffices to make the resulting system SOC. Another optimal solution is deploying a non-dedicated actuator that connects to $x_5$ and $x_4$ (or $x_5$ and $x_2$) simultaneously.} %It can be verified that, in all these cases, the resulting system is not structurally controllable.
\end{example}

      \begin{figure}
	\centering
	% Requires \usepackage{graphicx}
	\includegraphics[width=3.0in]{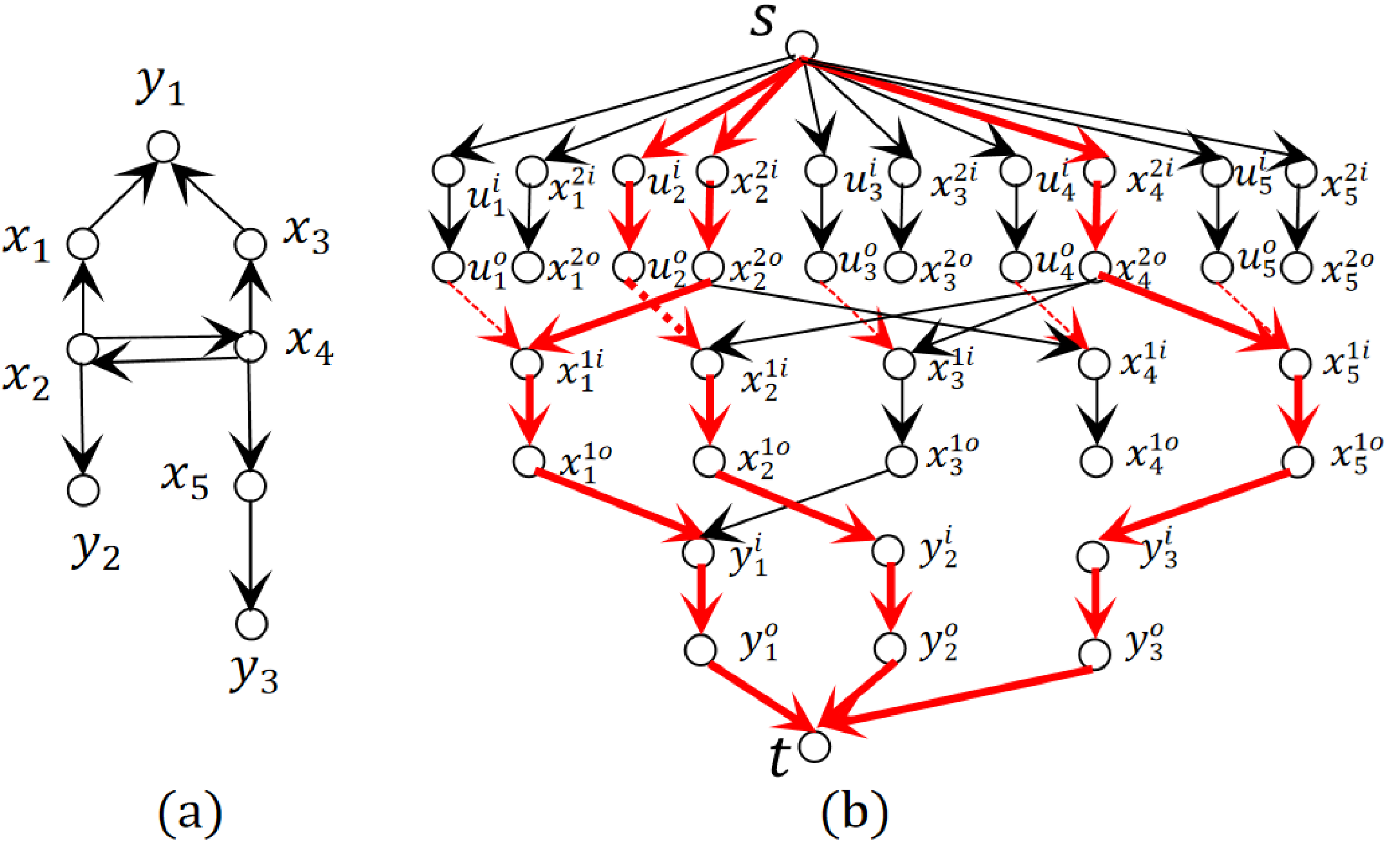}\\
	\caption{{Illustration of Algorithm \ref{alg4}. Subfigure (a) gives the digraph ${\mathcal G}(\bar A, \bar C)$. Subfigure (b) depicts the corresponding flow network ${\mathcal F}(\bar A, \bar C)$ constructed in step 1 of Algorithm \ref{alg4}. All edges have capacity one, dotted red edges have cost one, and the rest have cost zero. Bold red edges form a minimum cost maximum flow $f^*$ with total cost one, where each edge has flow $1$. Associated with it, $X^{f_1^*}=\{x_2\}$ and $X^{f_2^*}=\{x_2,x_4\}$.}} \label{driver-placement-soc}
\end{figure}

      \begin{remark}
      	Since adding additional links from actuators to states would not destroy SOC, if SOC can be verified in polynomial time, then a naive iterative algorithm like Algorithm \ref{alg2} can find an optimal solution to ${\mathcal P}_2$ for general systems. By contrast, even though SFO can be verified in polynomial time, the complexity of ${\mathcal P}_1$ remains open.
      \end{remark}

      {
      \begin{remark} The Matlab codes for Algorithms \ref{alg1}-\ref{alg4}, as well as for checking generic diagonalizability (Proposition \ref{check-diagonalizability}), are available at  https://github.com/Yuanzhang2014/Generic-diagonalizability-SFO-SOC.
      \end{remark}
      }
      	
      	\section{Conclusions} \label{sec-conclude}			
      	This paper has provided polynomial-time solutions to the verification and optimal sensor/actuator placement problems related to SFO and SOC in a class of systems, namely,   generically diagonalizable systems. We have defined generically diagonalizable matrices and provided the graph-theoretic characterizations. Computationally efficient criteria for SFO and SOC of generically diagonalizable systems were derived. For such systems, we provided closed-form solutions, as well as a weighted maximum matching-based algorithm and a weighted maximum flow-based algorithm, respectively, for determining the minimal sensors to achieve SFO and the minimal actuators to achieve SOC. For more general systems to achieve SFO, an upper bound was established by identifying a non-decreasing property of SFO w.r.t. a specific class of edge additions, which is shown to be optimal under certain conditions. 	 Future work includes exploring additional properties of generically diagonalizable systems, {characterizing the ratio of structurally diagonalizable networks within the total set of networks}, and investigating the complexity of the minimal sensor placement problem (${\mathcal P}_1$) in the general case. %Exploring more system properties of generically diagonalizable systems, characterizing the ratio of structurally diagonalizable networks within the total set of networks, and revealing the complexity of the minimal sensor placement problem (${\mathcal P}_1$) in the general case, are planned to be the future work. 	
      	
      	\section*{Acknowledgment}
      	The first author would like to thank Mr. Ranbo Cheng, Beijing Institute of Technology, for various discussions and help in constructing  Example \ref{example-counter}.
			
			% \section{Minimal sensor placement for  SFO} \label{sec-sensor}

				% It is straightforward from Lemma \ref{fundamental-theorem}	that (\ref{modal_functional}) is equivalent to ${\rm rank}[O(J_i,C_i);F_i]={\rm rank}O(J_i,C_i)$.

			}}
				
\section*{Appendix}
{\bf Proof of Proposition \ref{non-decreasing}:}
It suffices to consider the case ${\rm row}(\bar C)={\rm row}(\bar C')=p$, since adding more nonzero rows (i.e. sensors) to $\bar C'$ would not destroy the SFO provided that $(\bar A, \bar C', \bar F)$ is SFO. Without loss of generality, suppose $|\Delta E|\ge 1$. %We will prove by induction.
	Let $e\in \Delta E$ be an arbitrary edge of $\Delta E$. Suppose $e=(x_i,y_j)$, where $x_i\in X_F$ and $y_j\in Y\doteq \{y_1,...,y_p\}$. Let $\bar C^e$ be the output matrix obtained from $\bar C$ after adding $e$ to $E_{C}$.  We first show that $d({\mathcal G}(\bar A,\bar C^e))=d({\mathcal G}(\bar A, \bar C))$ by contradiction. If $d({\mathcal G}(\bar A,\bar C^e))>d({\mathcal G}(\bar A, \bar C))$, let
	${\mathcal G}_{\rm cact}$ be an arbitrary maximum cactus configuration in ${\mathcal G}(\bar A,\bar C^e)$. Since $(\bar A, \bar C, \bar F)$ is SFO, by \citet[Coro 4]{zhang2023functional}, every $x\in X_F$ is output-reachable in ${\mathcal G}(\bar A, \bar C)$, which means adding $e$ to ${\mathcal G}(\bar A, \bar C)$ will not affect the output-reachability of state vertices of ${\mathcal G}(\bar A, \bar C)$. As a result, $e$ must be contained in ${\mathcal G}_{\rm cact}$. Observe that by changing $e$ in ${\mathcal G}_{\rm cact}$ to $(x_i,{y_{p+1}})$, we can obtain a new cactus configuration of size $d({\mathcal G}(\bar A, \bar C^e))$ in ${\mathcal G}(\bar A, \bar C)\cup {\mathcal G}(x_i,y_a)$. Consequently, it holds that
	\begin{equation}\label{not_sfo_inequality} d({\mathcal G}(\bar A, \bar C)\cup {\mathcal G}(x_i,{y_{p+1}}))\ge d({\mathcal G}(\bar A, \bar C^e)).\end{equation}
	Inequality (\ref{not_sfo_inequality}) and the assumption $d({\mathcal G}(\bar A,\bar C^e))>d({\mathcal G}(\bar A, \bar C))$ yield $d({\mathcal G}(\bar A, \bar C)\cup {\mathcal G}(x_i,{y_{p+1}}))>d({\mathcal G}(\bar A, \bar C))$, which contracts the SFO of $(\bar A, \bar C, \bar F)$ from Lemma \ref{cactus-adding-criterion}. Therefore, it holds that $d({\mathcal G}(\bar A,\bar C^e))=d({\mathcal G}(\bar A, \bar C))$.
	
	We now prove that $(\bar A, \bar C^e, \bar F)$ is SFO by showing the SFO of $(\bar A, \bar C^e, \bar e_{i^*})$ for each $x_{i^*}\in X_F$. First, consider the case $i^*=i$. If $(\bar A, \bar C^e, \bar e_{i})$ is not SFO, Lemma \ref{cactus-adding-criterion} yields $d({\mathcal G}(\bar A, \bar C^e)\cup {\mathcal G}(x_i,{y_{p+1}}))> d({\mathcal G}(\bar A, \bar C^e)).$
	% \begin{equation}\label{not_sfo_inequality2} d({\mathcal G}(\bar A, \bar C^e)\cup {\mathcal G}(x_i,y_a))> d({\mathcal G}(\bar A, \bar C^e)),\end{equation} %where $y_a$ is an output vertex different from those in ${\mathcal G}(\bar A, \bar C^e)$.
	However, by the similar reasoning to (\ref{not_sfo_inequality}), it holds
	% $d({\mathcal G}(\bar A, \bar C'')\cup {\mathcal G}(x_i,y_a))= d({\mathcal G}(\bar A, \bar C)\cup {\mathcal G}(x_i,y_a)),$
	\begin{equation}\label{not_sfo_inequality3} d({\mathcal G}(\bar A, \bar C^e)\cup {\mathcal G}(x_i,{y_{p+1}}))= d({\mathcal G}(\bar A, \bar C)\cup {\mathcal G}(x_i,{y_{p+1}})),\end{equation}
	which comes from the fact that every state vertex must have a unique outgoing edge in a cactus configuration of ${\mathcal G}(\bar A, \bar C^e)\cup {\mathcal G}(x_i,{y_{p+1}})$, and each state vertex $x\in X_F$ has already been output-reachable in ${\mathcal G}(\bar A, \bar C)$. These relations together yield  $d({\mathcal G}(\bar A, \bar C)\cup {\mathcal G}(x_i,{y_{p+1}}))>d({\mathcal G}(\bar A, \bar C))$, contradicting the SFO of $(\bar A, \bar C, \bar e_i)$. Therefore, $(\bar A, \bar C^e, \bar e_{i})$ is SFO.
	Next, consider the case $i^*\ne i$. Let ${y_{p+2}}$ be the $(p+2)$th output vertex. By the similar reasoning to (\ref{not_sfo_inequality}) and (\ref{not_sfo_inequality3}), we have $d({\mathcal G}(\bar A, \bar C^e)\cup {\mathcal G}(x_{i^*},{y_{p+2}}))\le d({\mathcal G}(\bar A, \bar C)\cup {\mathcal G}(x_i,{y_{p+1}})\cup {\mathcal G}(x_{i^*},{y_{p+2}})).$
	% \begin{equation}\label{not_sfo_inequality4} d({\mathcal G}(\bar A, \bar C^e)\cup {\mathcal G}(x_{i^*},y_b))\le d({\mathcal G}(\bar A, \bar C)\cup {\mathcal G}(x_i,y_a)\cup {\mathcal G}(x_{i^*},y_b)).\end{equation}
	On the other hand, from condition (d) of Corollary \ref{single-pro}, the SFO of $(\bar A, \bar C, \bar F)$ requires that $(\bar A, \bar C, \bar F')$ is SFO with $X_{F'}=\{x_i,x_{i^*}\}$. This fact with Lemma \ref{cactus-adding-criterion} yields $d({\mathcal G}(\bar A, \bar C)\cup {\mathcal G}(x_i,{y_{p+1}})\cup {\mathcal G}(x_{i^*},{y_{p+2}}))=d({\mathcal G}(\bar A, \bar C)).$
	% \begin{equation}\label{not_sfo_inequality5} d({\mathcal G}(\bar A, \bar C)\cup {\mathcal G}(x_i,y_a)\cup {\mathcal G}(x_{i^*},y_b))=d({\mathcal G}(\bar A, \bar C)).\end{equation}
	Since $d({\mathcal G}(\bar A, \bar C^e)\cup {\mathcal G}(x_{i^*},{y_{p+2}}))\ge d({\mathcal G}(\bar A, \bar C^e))$, we reach $d({\mathcal G}(\bar A, \bar C^e)\cup {\mathcal G}(x_{i^*},{y_{p+2}}))=d({\mathcal G}(\bar A, \bar C))$,
	% \begin{equation}\label{not_sfo_inequality6} d({\mathcal G}(\bar A, \bar C^e)\cup {\mathcal G}(x_{i^*},y_b))=d({\mathcal G}(\bar A'', \bar C)),\end{equation}
	which implies that $(\bar A, \bar C^e, \bar e_{i^*})$ is SFO. Since this holds for each $x_{i^*}\in X_F$, $(\bar A, \bar C^e, \bar F)$ is SFO via Corollary \ref{single-pro}.

	Let $\bar C^{e'}$ be the output matrix corresponding to $\{e'\}\cup E_{\bar C^e}$ for any $e'\in \Delta E \backslash \{e\}$.
	Similarly to the above analysis, we can prove that $(\bar A, \bar C^{e'}, \bar F)$ is SFO from the SFO of $(\bar A, \bar C^{e}, \bar F)$. Repeat this procedure $|\Delta E|$ times, and we obtain that $(\bar A, \bar C', \bar F)$ is SFO.  {\hfill $\square$\par}
	
	{\bf Proof of Theorem \ref{theorem-general-case}:}
	  %    \begin{proof}
		The feasibility of Algorithm \ref{alg2} follows from the non-decreasing property of SFO w.r.t. functional state measurement addition as analyzed above. Below, we focus on the statements about Algorithm \ref{alg3}. First, observe that given a collection of disjoint cycles and stems in ${\mathcal G}(\bar A, \bar I_{X_F})$ (note that vertices in the cycles need not be output-reachable), we can complete each stem to a cycle by adding the zero cost edge from its terminal vertex in $Y$ to its initial vertex in $X$, and then add the loops for uncovered vertices in $X\cup Y$. Then $X\cup Y$ are covered by disjoint cycles, corresponding to a maximum (perfect) matching of $\tilde {\mathcal B}(\bar A, \bar I_{X_F})$. Conversely, given a maximum matching in $\tilde {\mathcal B}(\bar A, \bar I_{X_F})$, by omitting the zero cost edges, we obtain a collection of disjoint cycles and stems (as well as isolated vertices) in ${\mathcal G}(\bar A, \bar I_{X_F})$.
		Consider a maximum matching ${\mathcal H}$ of $\tilde {\mathcal B}(\bar A, \bar I_{X_F})$ and let the indicator function ${\mathbb I}:\tilde E_{XY}\to \{0,1\}$ be such that ${\mathbb I}({\mathcal H},e)=1$ if $e\in {\mathcal H}$ and ${\mathbb I}({\mathcal H},e)=0$ otherwise.
		Observe that the weight $w({\mathcal H})$ of a maximum matching ${\mathcal H}$ in $\tilde {\mathcal B}(\bar A, \bar I_{X_F})$ can be equivalently written as
		$$w({\mathcal H})\doteq (q+1)\sum_{e\in E_{XY}^F\doteq E_{XX}^F\cup E_{XY}}{\mathbb I}({\mathcal H},e)- \sum_{e\in E_{XY}}{\mathbb I}({\mathcal H},e).$$
		Suppose ${\mathcal H}^*$ is a maximum matching with the maximum weight in $\tilde {\mathcal B}(\bar A, \bar I_{X_F})$. Then, it must hold
		$\sum_{e\in E_{XY}^F}{\mathbb I}({\mathcal H}^*,e)={\rm grank}\,O(\bar A, \bar I_{X_F})$. To show this, assume, for the sake of contradiction, $\sum_{e\in E_{XY}^F}{\mathbb I}({\mathcal H}^*,e)<{\rm grank}\,O(\bar A, \bar I_{X_F})$ (noting that all vertices in the set $X\backslash W_F$ are output-unreachable, by Lemma \ref{hosoe-subspace}, we have $\sum_{e\in E_{XY}^F}{\mathbb I}({\mathcal H},e)\le d({\mathcal G}(\bar A, \bar I_{X_F}))={\rm grank}\,O(\bar A, \bar I_{X_F})$ for any maximum matching ${\mathcal H}$ of $\tilde {\mathcal B}(\bar A, \bar I_{X_F})$).  Consider a maximum matching ${\mathcal H}'$ in $\tilde {\mathcal B}(\bar A,\bar I_{X_F})$ such that $\sum_{e\in E_{XY}^F}{\mathbb I}({\mathcal H}',e)={\rm grank}\,O(\bar A, \bar I_{X_F})$ (such a maximum matching always exists by Lemma \ref{hosoe-subspace}).  Observe that
		\begin{align*}
		&w({\mathcal H}^*)-w({\mathcal H}')=(q+1)(\sum_{e\in E_{XY}^F}{\mathbb I}({\mathcal H}^*,e)\\&-\sum_{e\in E_{XY}^F}{\mathbb I}({\mathcal H}',e))-(\sum_{e\in E_{XY}}{\mathbb I}({\mathcal H}^*,e)-\sum_{e\in E_{XY}}{\mathbb I}({\mathcal H}',e))\\ & \le
		-(q+1)+q \\&< 0,
		\end{align*}where the first inequality results from the fact that $0\le \sum_{e\in E_{XY}}{\mathbb I}({\mathcal H}',e), \sum_{e\in E_{XY}}{\mathbb I}({\mathcal H}^*,e) \le q$. This contradicts the assumption that ${\mathcal H}^*$ is a maximum weighted maximum matching. As a result of the above analysis, ${\mathcal H}^*$ has the minimum $\sum_{e\in E_{XY}}{\mathbb I}({\mathcal H}^*,e)$ over all perfect matchings ${\mathcal H}$ in $\tilde {\mathcal B}(\bar A, \bar I_{X_F})$ satisfying $\sum_{e\in E_{XY}^F}{\mathbb I}({\mathcal H},e)={\rm grank}\,O(\bar A, \bar I_{X_F})$. Equivalently,
		${\mathcal H}^*$ corresponds to a maximum cactus configuration of ${\mathcal G}(\bar A, \bar I_{X_F})$ with the minimum number of output stems (this number equals $\sum_{e\in E_{XY}}{\mathbb I}({\mathcal H}^*,e)$). It turns out that $\gamma\doteq \max\{1,\sum_{e\in E_{XY}}{\mathbb I}({\mathcal H}^*,e)\}$ is the minimum number of rows of an output matrix $\bar C$ subject to $X_{C}\subseteq X_F$ such that
		${\rm grank}\,O(\bar A, \bar C)={\rm grank}\,O(\bar A, \bar I_{X_F})$. Indeed, if there is another $\bar C'$ with fewer rows than $\bar C$ that satisfies the aforementioned condition, then a maximum cactus configuration in ${\mathcal G}(\bar A, \bar C')$ exists with fewer output stems than $\sum_{e\in E_{XY}}{\mathbb I}({\mathcal H}^*,e)$, which is also a maximum cactus configuration in ${\mathcal G}(\bar A, \bar I_{X_F})$, causing a contradiction.
		Now consider the output matrix $\bar C$ found by Algorithm \ref{alg3}. By the construction, any state vertex that is output-reachable in ${\mathcal G}(\bar A, \bar I_{X_F})$ is also output-reachable in ${\mathcal G}(\bar A, \bar C)$. Consequently, ${\rm grank}\,O(\bar A,\bar C)=\sum_{e\in E_{XY}^F}{\mathbb I}({\mathcal H}^*,e)={\rm grank}\,O(\bar A,\bar I_{X_F})$. Since $X_C\subseteq X_F$, it holds that ${\rm grank}\,O(\bar A,[\bar C;\bar I_{X_F}])={\rm grank}\,O(\bar A,\bar I_{X_F})$, leading to ${\rm grank}\,O(\bar A,\bar C)={\rm grank}\,O(\bar A,[\bar C;\bar I_{X_F}])$. By Corollary \ref{single-pro}, $(\bar A, \bar C, \bar F)$ is SFO.
		
		The optimality when $\bar F=\bar I_n$ is obvious from the above analysis. Since the optimal solution $\bar C$ returned by Algorithm \ref{alg1} satisfies $X_{C}\subseteq X_F$, the solution given by Algorithm \ref{alg2} or \ref{alg3} is also optimal for generically diagonalizable systems.	 {\hfill $\square$\par}
%	\end{proof}

				%			\bibliographystyle{elsarticle-num}
				%			{\footnotesize
					%				\bibliography{yuanz3}
					%			}
				
				%		\section*{\refname}
				%\bibliographystyle{elsarticle-num}
     \bibliographystyle{elsarticle-harv}
				{\small
					\bibliography{yuanz3}
				}
			\end{document}